\documentclass[12pt,twoside,a4paper]{article}
\usepackage{a4wide}
\usepackage{epsfig}
\usepackage{rotating}
\usepackage{cite}
\usepackage{cerntit}

\date{December 19, 2000}

\def\pz{\phantom{0}}
\def\pzz{\phantom{00}}

\newcommand{\mth}      {\multicolumn {3} {|c}}

\newcommand{\Lep}   {\textsc{Lep}}
\newcommand{\Aleph}  {\textsc{Aleph}}
\newcommand{\Delphi} {\textsc{Delphi}}
\newcommand{\Lthree} {\textsc{L3}}
\newcommand{\Opal}   {\textsc{Opal}}

\newcommand{\eV}{\ensuremath{\mathrm{e\kern -0.12em V}}}
\newcommand{\MeV}{\ensuremath{\mathrm{Me\kern -0.12em V}}}
\newcommand{\GeV}{\ensuremath{\mathrm{Ge\kern -0.12em V}}}
\newcommand{\TeV}{\ensuremath{\mathrm{Te\kern -0.12em V}}}

\newcommand{\MZ}{\ensuremath{m_\mathrm{Z}}}
\newcommand{\GZ}{\ensuremath{\Gamma_\mathrm{Z}}}
\newcommand{\MZA}{\ensuremath{m_\mathrm{Z}^{90-92}}}

\newcommand{\MZB}{\ensuremath{m_\mathrm{Z}^{93-94}}}

\newcommand{\MZC}{\ensuremath{m_\mathrm{Z}^{95}}}

\newcommand{\Zzero}   {\mbox{${\mathrm{Z}}$}}

\newcommand {\GF}      {G_{\mathrm{F}}}
\newcommand {\MH}      {m_{\mathrm{H }}}
\newcommand {\Mt}      {m_{\mathrm{t}}}
\newcommand {\shad}  {\sigma_{\rm h}^0}
\newcommand {\slept}  {\sigma_{\rm \ell}^o}
\newcommand {\Ree}      {R_{\mathrm{e}}}
\newcommand {\Rmu}      {R_{\mu}}
\newcommand {\Rtau}      {R_{\tau}}
\newcommand {\Rl}      {R_{\ell}}
\newcommand{\Afb}     {A_{\mathrm{fb}}}
\newcommand {\Afbpol}     {\rm{A}\!^{0,\,\ell}_{\rm {FB}}}
\newcommand {\Afbze}     {\rm{A}\!^{0,\,{\rm e}}_{\rm {FB}}}
\newcommand {\Afbzf}     {\rm{A}\!^{0,\,{\rm f}}_{\rm {FB}}}
\newcommand {\Afbzm}     {\rm{A}\!^{0,\,\mu}_{\rm {FB}}}
\newcommand {\Afbzt}     {\rm{A}\!^{0,\,\tau}_{\rm {FB}}}

\newcommand {\Gff}        {\Gamma_{\rm {ff}}}
\newcommand {\Ginv}       {\Gamma_{\mathrm{inv}}}
\newcommand {\Ginvx}       {\Gamma^x_{\mathrm{inv}}}
\newcommand {\Ghad}       {\Gamma_{\mathrm{had}}}
\newcommand {\Gl}        {\Gamma_{\ell\ell}}
\newcommand {\Gnu}        {\Gamma_{\nu\nu}}
\newcommand {\Gee}        {\Gamma_{\rm {ee}}}
\newcommand {\Gmumu}      {\Gamma_{\mu\mu}}
\newcommand {\Gtautau}    {\Gamma_{\tau\tau}}

\newcommand {\cAf} {\mbox{${\cal A}_{\rm f}$}}
\newcommand {\cAe} {\mbox{${\cal A}_{\rm e}$}}

\newcommand{\ga}{g_{A}}
\newcommand{\gv}{g_{V}}
\newcommand{\gnu}{g^\nu}

\newcommand{\gve}{g_{V}^{\rm e}}
\newcommand{\gaf}{g_{A}^{\rm f}}
\newcommand{\gvf}{g_{V}^{\rm f}}

\newcommand{\gvq}{g_{V}^{\rm q}}

\newcommand{\gal}{g_{A}^{\ell}}
\newcommand{\gvl}{g_{V}^{\ell}}

\newcommand{\cga}{{\cal G}_{A}}
\newcommand{\cgv}{{\cal G}_{V}}
\newcommand{\cgae}{{\cal G}_{A}^{\rm e}}
\newcommand{\cgve}{{\cal G}_{V}^{\rm e}}
\newcommand{\cgaf}{{\cal G}_{A}^{\rm f}}
\newcommand{\cgvf}{{\cal G}_{V}^{\rm f}}
\newcommand{\Raf}{R_{A}^{\rm f}}
\newcommand{\Rvf}{R_{V}^{\rm f}}

\newcommand {\rhoeffl}    {\rho_{\rm eff}^{\rm {lept}}}
\newcommand {\rhoeffnu}    {\rho_{\rm eff}^{\nu}}

\newcommand {\swsqeff}    {\sin^2\!\theta_{\rm eff}}
\newcommand {\swsqeffl}    {\sin^2\!\theta_{\rm eff}^{\rm {lept}}}

\newcommand {\cost}       {\cos\theta}

\newcommand{\ff}{{\rm f}\overline{\rm f}}
\newcommand{\qq}{{\rm q}\overline{\rm q}}
\newcommand{\lept}{\ell^+\ell^-}
\newcommand{\ee}{\mathrm{e}^+\mathrm{e}^-}

\newcommand{\mumu}{\mu^+\mu^-}

\newcommand{\tautau}{\tau^+\tau^-}
\newcommand{\eeff}{\ee\rightarrow\ff}

\newcommand{\eeee}{\ee\rightarrow\ee}

\newcommand{\eemumu}{\ee\rightarrow \mu^+\mu^-}
\newcommand{\eetautau}{\ee\rightarrow \tau^+\tau^-}

\newcommand {\Ztonunu}   {\mathrm{Z}\rightarrow \nu\overline{\nu}}

\begin{document}
                             \preprint{CERN-EP-2000-153 }    
\begin{titlepage}

  \title{\bf
      Combination procedure for the
     precise determination \\
      of Z boson parameters from \\
    results of the LEP experiments }

\vspace*{-6pc}
\begin{center}
{\large The \Lep~Collaborations  
 \Aleph, \Delphi, \Lthree~and {\Opal}\,\footnote{The full 
list of authors may be found in 
References~\citen{ALEPHLS,DELPHILS,L3LS,OPALLS}}
    and \\
  the Line Shape Sub-group of the {\Lep} Electroweak Working Group}\,\footnote{ 
The members of the line shape group are: 
                   G.~Duckeck, 
                   M.~Gr\"{u}newald,
                   T.~Kawamoto,
                   R.~Kellogg, 
                   G.~Martinez,
                   J.~Mnich,
                   A.~Olshevski,
                   B.~Pietrzyk, 
                   G.~Quast, 
                   P.~Renton,
                   E.~Tournefier } 
\end{center}

 \begin{abstract}
    The precise determination of the Z boson parameters from the 
    measurements at the Z resonance by the four collaborations \Aleph, 
    {\Delphi}, {\Lthree} and {\Opal} in $\ee$ collisions at the large electron
    positron collider {\Lep} at CERN is a landmark for precision tests
    of the electroweak theory. The four experiments measured quantities
    which were used to extract the
    mass and width of the Z boson, the hadronic cross-section at 
    the pole of the resonance, the ratio of hadronic 
    and leptonic decay widths, and the leptonic forward-backward 
    asymmetries at the pole. The combination procedure based on 
    these parameters is presented in this paper. \\
~\\
      \hspace*{5em}(to be published as part of a review in Physics Reports)
 \end{abstract} 

\end{titlepage}


\newpage
\section{Introduction}

Between the years 1989 and 1995 the $\mathrm{e^+e^-}$ collider {\Lep} at
CERN provided interactions at centre-of-mass energies, $\sqrt{s}$, 
ranging from 88 to 95~$\GeV$, {\sl i.\,e.} around the mass of the 
$\Zzero$ boson (\Lep\,I phase).  An important aspect of
physics at {\Lep} concerns the analysis of fermion-pair production in
$\mathrm{e^+e^-}$ collisions~\cite{LEP1-yellow}.  The four {\Lep}
experiments {\Aleph}~\cite{ALEPHdet}, {\Delphi}~\cite{DELPHIdet}, 
{\Lthree}~\cite{L3det} and {\Opal}~\cite{OPALdet} analysed, in particular, 
hadron (quark-pair) production and the pair production of charged 
leptons, $\ell={\rm e},\mu,\tau$.

At various centre-of-mass energies, total cross-sections are measured
for all processes, while forward-backward asymmetries are
measured in lepton-pair production.
These measurements (``{\it realistic observables}'') allow the 
determination of various properties of the $\Zzero$
boson such as its mass, total and partial decay widths, and coupling
constants to fermions (``{\it pseudo-observables}'').  
For the extraction of the pseudo-observables, the ex\-pe\-ri\-ments 
perform model-independent fits to their measured realistic 
observables~\cite{ALEPHLS,DELPHILS,L3LS,OPALLS}.

To obtain the best possible precision the results of the four 
{\Lep} experiments have to be averaged. This paper describes the 
combination procedure adopted by the {\Lep} electroweak working 
group.  Performing an average over the realistic 
observables constitutes an extremely complicated task, as it involves 
hundreds of measurements, each with specific phase space definitions
and experimental errors which are correlated among different 
centre-of-mass energies and data
taking periods and also among the {\Lep} experiments. Therefore, the
combination of the experimental results is performed on the basis of the
four sets of pseudo-ob\-ser\-vab\-les. As will be shown here, this is
possible without significant loss of precision.

Additional fits to the experimental data, usually not contained 
in the individual publications quoted above, were provided by 
the experiments and are documented in this paper. A large effort 
is devoted to the treatment of systematic errors and their correlation 
among the experiments, such that the combination procedure 
yields an optimal estimator for the averages.

This paper is organised as follows: 
Section\,\ref{sec-datataking} summarises the information about the {\Lep}~I 
running relevant for this paper.
Section\,\ref{sec-zpar} presents a brief introduction to the 
pseu\-do-ob\-ser\-vab\-les
used to parametrise the realistic observables around the $\Zzero$ 
resonance. The individual experimental results are presented in 
Section\,\ref{sec-inpdata}. Sources of correlated systematic errors 
between experiments and their effects on the pseudo-observables 
are discussed in Section\,\ref{sec-comerr}.
In Section\,\ref{sec-combine}, studies of various methods for 
combining the results are presented. The resulting 
pseudo-observables are then considered in the framework of a 
specific model, the minimal Standard Model (SM), and are compared
with direct SM fits to the realistic observables.
The conclusions are summarised in Section\,\ref{sec-conclusion}.

\section{Z resonance scans at LEP\,I \label{sec-datataking}}

\begin{figure}[htb]
\begin{center}
\mbox{\epsfig{file=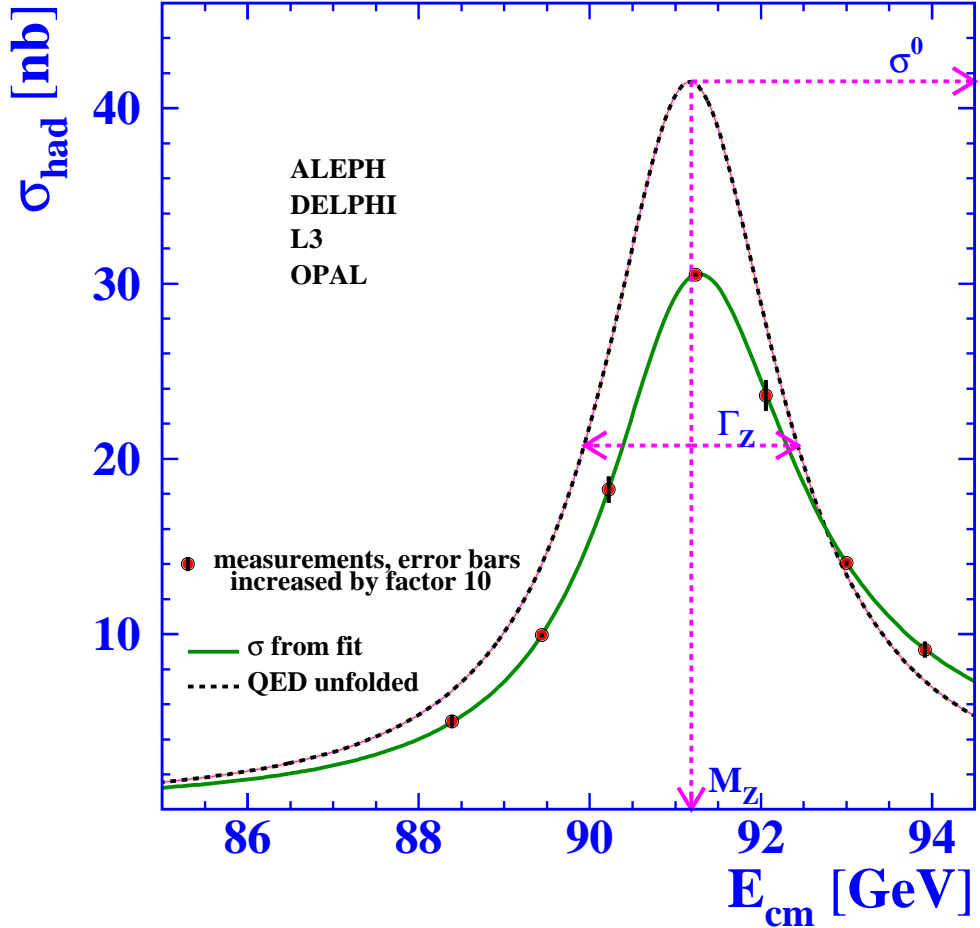,width=0.8\textwidth}} 
\end{center}
\caption[Average hadronic cross-section measurements]
{\em \label{fig:xsh_all} Average over measurements 
of the hadronic cross-sections by the four experiments, as
a function of centre-of-mass energy. The dashed curve shows 
the QED de-convoluted cross-section, which defines the $\Zzero$ 
parameters described in the text.} 
\end{figure}

Running of {\Lep} in the years from 1989 to 1995 was dedicated to precision 
studies of the $\Zzero$ boson parameters. Electron-positron collisions 
were provided at several well-determined centre-of-mass energies around 
the $\Zzero$ resonance, with steadily improving performance. The set 
of measurements collected by the experiments consists 
of the hadronic and leptonic cross-sections and the leptonic 
forward-backward asymmetries around seven points in centre-of-mass 
energy, over six years of running at \Lep\,I. In addition, changes of 
experimental conditions, such as the inclusion of new detector components, 
made it necessary to subdivide the data samples even further. 
The full {\Lep}~I data set consists of about $4 \times 200$ 
individual cross-section and asymmetry measurements. From these each 
experiment has extracted a set of parameters describing the 
cross-section around the $\Zzero$ resonance,
which include the mass, $\MZ$, and width, $\GZ$, of the $\Zzero$ and the
total pole cross-section for $\qq$ production, $\shad$. These parameters 
are discussed in detail in Section\,\ref{sec-zpar}\,. QED corrections from 
initial-state photon radiation are large around the 
$\Zzero$ resonance due to the rapid variation 
of the cross-sections with centre-of-mass energy. For illustration, 
Figure\,\ref{fig:xsh_all} shows the average over the hadronic 
cross-section measurements by the  four experiments, together 
with the fitted line shape curve before and after unfolding 
photon radiation. The cross-section is dominated
by on-shell $\Zzero$ production, although photon exchange and 
$\gamma$-$\Zzero$ interference contributions are not negligible.
The measurements are sensitive to higher-order electroweak corrections.
These modify the tree-level couplings of the $\Zzero$ to fermions,
and are quantified in terms of electroweak form factors.

Much effort was dedicated to the determination of the energy of the 
colliding beams, which reached 
a precision of about $20\times10^{-6}$ on the absolute energy scale. 
This level of accuracy is vital 
for the measurements of the mass and width of the 
$\Zzero$. All the experiments replaced their first-generation luminosity 
detectors, which had systematic uncertainties around the percent level, 
by high-precision devices capable of pushing systematic errors on the
acceptance of small-angle Bhabha scattering events below one per-mil. 
As a consequence of improvements of the accelerator and of 
the experiments during {\Lep}~I running, 
statistical and systematic errors are much smaller for the 
last three years of data taking, which hence dominate the precision 
achieved on the $\Zzero$ parameters.

\subsection{Event selection and statistics \label{sec-evstat}}

During the  summer of 1989 the first $\Zzero$ bosons were produced at {\Lep} 
and observed by the four experiments. Since then the operation of the machine 
and its performance were steadily improved. At the end of data taking 
around the $\Zzero$ resonance in autumn 1995 the peak luminosity 
had reached nearly twice its design value. Table\,\ref{tab:lepop} summarises 
the data taking periods, the approximate centre-of-mass energies and the
integrated luminosities delivered. 

\begin{table}[tbph]
 \begin{center}
  \begin{tabular}{c|cr}
 year  & beam energy range & integrated   \\
       &   ~[$\GeV$]       & ~luminosity  \\
       &                 & ~[pb$^{-1}$] \\
\hline 
 1989  & [88.2\,,\,94.2] &  1.7        \\
 1990  & [88.2\,,\,94.2] &  8.6        \\
 1991  & [88.5\,,\,93.7] & 18.9        \\
 1992  & 91.3            & 28.6        \\
 1993  & 89.4,\,91.2,\,93.0 & 40.0        \\
 1994  & 91.2            & 64.5        \\
 1995  & 89.4,\,91.3,\,93.0 & 39.8        \\
\end{tabular}
\end{center}
\caption[Beam energies and integrated luminosities]
{\em \label{tab:lepop} Approximate centre-of-mass energies and 
   integrated luminosities delivered by {\Lep}, per experiment. 
In 1990 and 1991, a total of about 7~pb$^{-1}$
was taken at off-peak energies, and 20~pb$^{-1}$ per year in 1993 
and in 1995. The total luminosity used by the experiments in
the analyses was smaller by 10-15\,\% due to data taking inefficiencies. }
\end{table}
The data collected in 1989 constitute only a very small subset of the total
statistics and are of lower quality, and therefore are not used here. 
In the years 1990 and 1991 ``energy scans'' were performed at seven 
different centre-of-mass energies around the peak of the $\Zzero$ 
resonance, placed about one~$\GeV$ apart. In 1992 and 1994 there were
high-statistics  runs at the peak energy only. In 1993 and 1995 data 
taking took place at three energy points, about $1.8\,\GeV$ below and 
above the peak and at the peak. In particular the off-peak energies 
were carefully calibrated employing the technique of resonant 
depolarisation of the transversely polarised 
beams\,\cite{bib-MZpaper,bib-ECAL92,bib-ECAL93,bib-ECAL95}. 

The accumulated event statistics amount to about 17 million 
$\Zzero$ decays recorded by the four experiments. 
A detailed breakdown is given in Table\,\ref{tab:LSstat}. 
\begin{table}[hbtp]
\begin{center}\begin{tabular}{lr} 
\begin{minipage}[b]{0.49\textwidth}
\begin{center}\begin{tabular}{r||rrrr||r}
  \multicolumn{6}{c}{$\qq$}  \\
\hline
   year & A &   D  &   L  &  O  & all \\
\hline
'90/91  & 433 &  357 &  416 &  454 &  1660\\
'92     & 633 &  697 &  678 &  733 &  2741\\
'93     & 630 &  682 &  646 &  649 &  2607\\
'94     &1640 & 1310 & 1359 & 1601 &  5910\\
'95     & 735 &  659 &  526 &  659 &  2579\\
\hline
 total  & 4071 & 3705 & 3625 & 4096 & 15497\\
\end{tabular}\end{center}
\end{minipage} 
   &
\begin{minipage}[b]{0.49\textwidth}
\begin{center}\begin{tabular}{r||rrrr||r}
  \multicolumn{6}{c}{$\lept$} \\
\hline
   year & A &   D  &   L  &  O  & all \\
\hline
'90/91  &  53 &  36 &  39  &  58  &  186 \\
'92     &  77 &  70 &  59  &  88  &  294 \\
'93     &  78 &  75 &  64  &  79  &  296 \\
'94     & 202 & 137 & 127  & 191  &  657 \\
'95     &  90 &  66 &  54  &  81  &  291 \\
\hline
total   & 500 & 384 & 343  & 497  & 1724 \\
\end{tabular}\end{center}
\end{minipage} \\
\end{tabular} \end{center}
\caption[Recorded event statistics]{\em \label{tab:LSstat}
The $\qq$ and $\lept$ event statistics, in units of $10^3$, used 
for the analysis of the $\Zzero$ line shape and lepton forward-backward
asymmetries by the experiments {\Aleph} (A), {\Delphi} (D), {\Lthree} (L)  
and {\Opal} (O).}
\end{table}

\begin{figure}[tbp]
\begin{sideways}
\begin{minipage}{\textheight}
\mbox{\epsfig{file=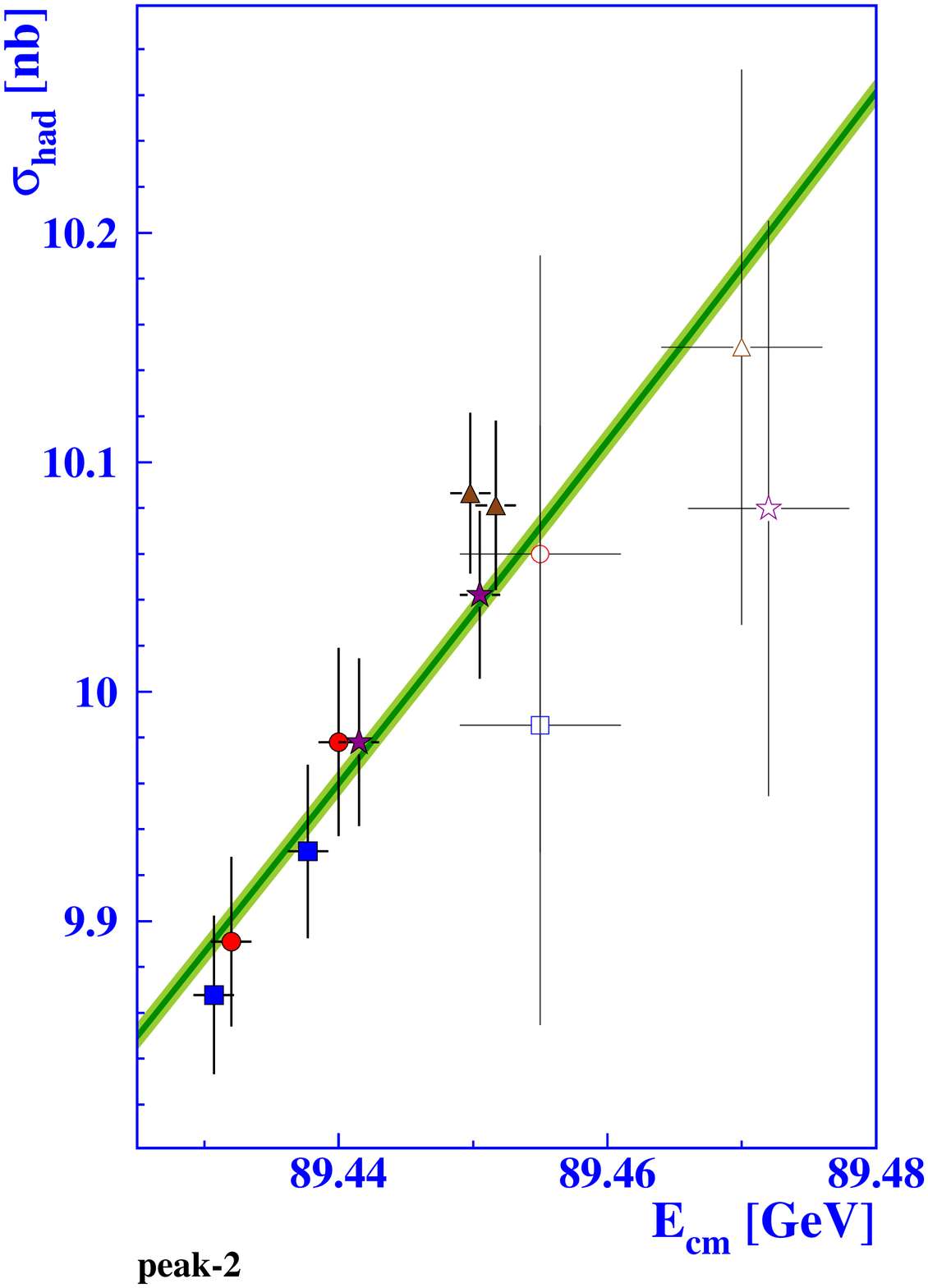,width=0.333\textheight}} 
\mbox{\epsfig{file=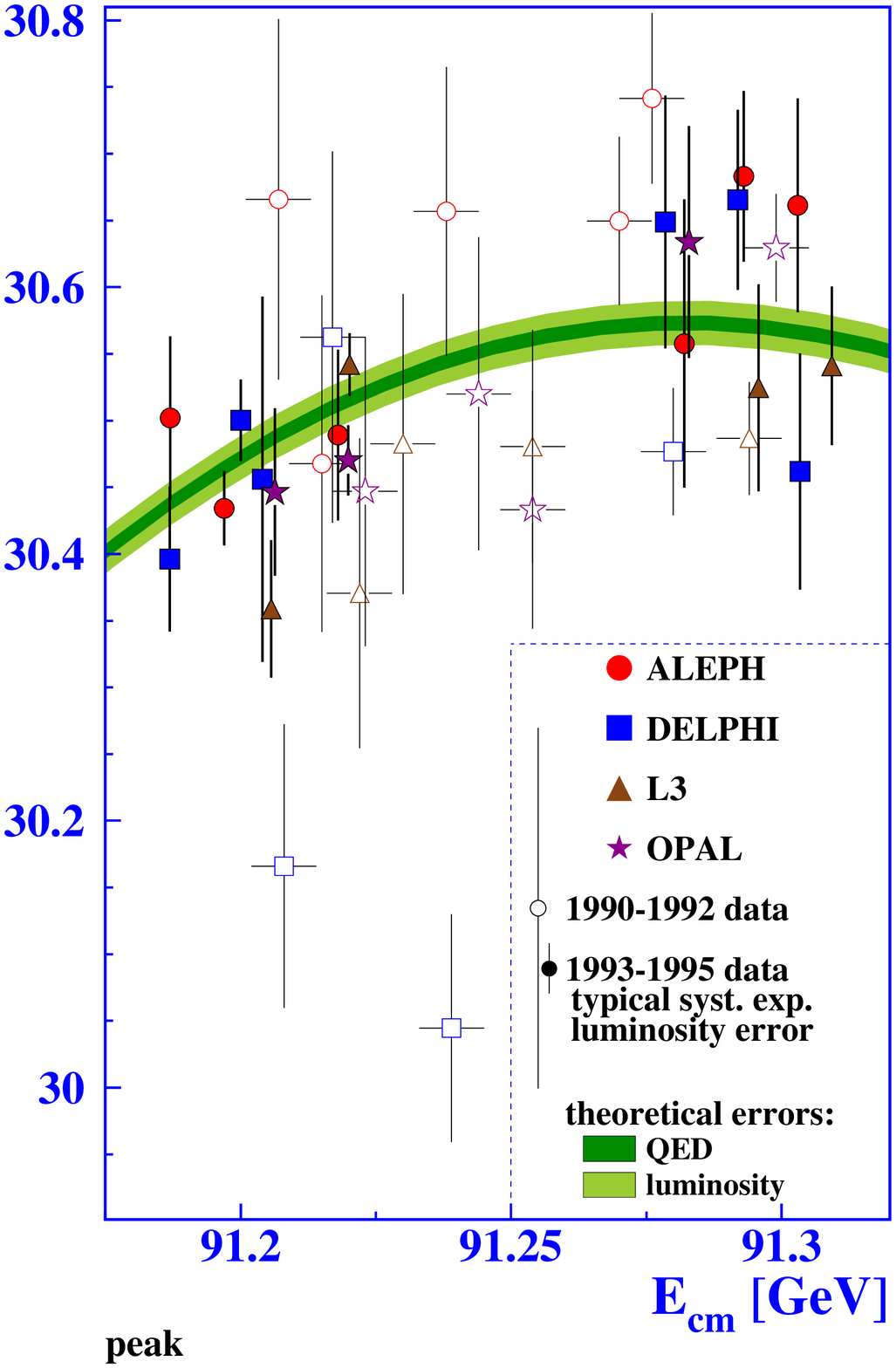,width=0.333\textheight}} 
\mbox{\epsfig{file=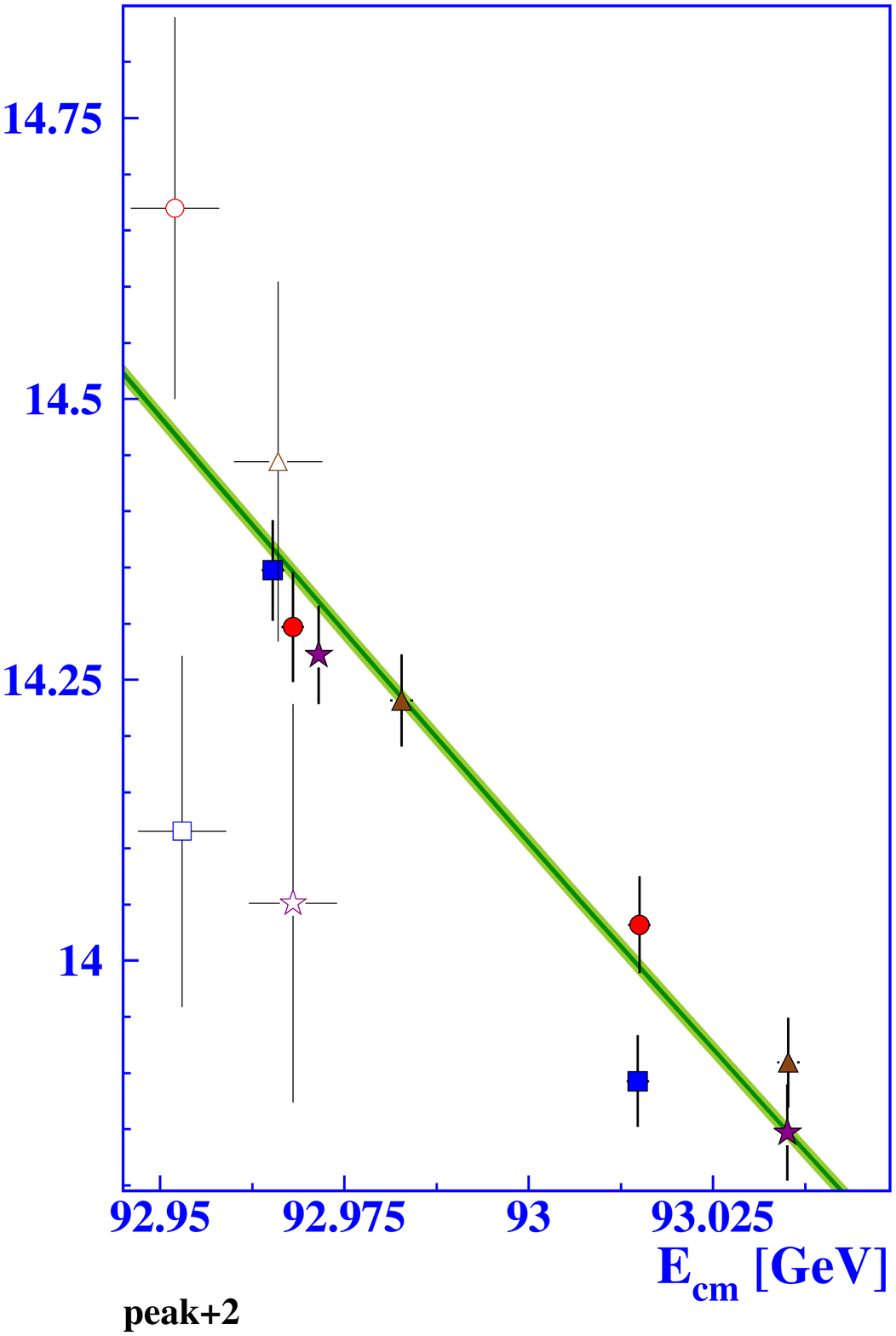,width=0.333\textheight}} \\
\caption[Hadronic cross-section measurements]
{\em \label{fig:xsh} Measurements by the four experiments of the 
hadronic cross-sections around the three principal energies.
The vertical error bars show the statistical errors only. The 
open symbols represent the early measurements with typically much 
larger systematic errors than the later ones, shown as
full symbols. Typical experimental systematic errors on the 
determination of the luminosity are indicated in the legend; these are
almost fully correlated within each experiment, but uncorrelated 
among the experiments.  The horizontal error bars show the 
uncertainties in {\Lep} centre-of-mass energy, where the errors
for the period 1993--1995 are smaller than the symbol size in 
some cases. The bands represent the result of the model-independent 
fit to all data, including the two most important common
theoretical errors from initial-state photon radiation and from 
the calculations of the small-angle Bhabha cross-section.}
\end{minipage}
\end{sideways}
\end{figure}
As an example, the measurements of the hadronic cross-section 
at the three principal energy points are shown in 
Figure\,\ref{fig:xsh}. Because the hadron
statistics are almost ten times larger than the lepton statistics, 
these measurements dominate the determination of the mass and the 
width of the $\Zzero$. Detailed descriptions of the individual 
experimental analyses can be found 
in~\cite{ALEPHLS,DELPHILS,L3LS,OPALLS}. They all rely on excellent
separation of the final states, $\qq$, $\ee$, $\mumu$ and $\tautau$,
accompanied by high selection efficiencies. The total cross-section,
$\sigma_{\rm tot}$, 
is determined from the number of selected events in a final state, 
$N$, the number of background
events, $N_{\rm bg}$, the selection efficiency, $\epsilon$, and
the integrated luminosity, ${\cal L}$, as
\( \sigma_{\rm tot}=(N-N_{\rm bg})/(\epsilon {\cal L} )~. \)

The luminosity of the beams is determined by normalisation to the
theoretical cross-section for the process of small-angle Bhabha 
scattering, which is dominated by 
photon exchange in the $t$ channel. Thus, the integrated
luminosity is given by the number of observed 
small-angle Bhabha events 
and the calculated cross-section for this process
within a given experimental acceptance. 
This requires the detection
of back-to-back electrons and positrons close to the beam direction. 
Their positions and energies are precisely measured by
forward calorimeters placed at small angles with  respect to
the beam line (typically 30~mrad$\,<\theta<$\,50~mrad).
The main experimental systematic error arises from the definition of
the geometrical acceptance for this process. Since the angular
distribution is steeply falling with increasing scattering angle
($\propto\theta^{-3}$), the precise definition of the
inner radius of the acceptance region is most critical. 

The forward-backward asymmetry, $\Afb$, is defined by the numbers of
events in which the final state lepton goes forward or backward
with respect to the direction of the incoming electron,
$N_{\rm f}$ and $N_{\rm b}$, respectively:
\(\Afb=(N_{\rm f}-N_{\rm b})/(N_{\rm f}+N_{\rm b})~.\)
In practice, $\Afb$ is determined from
a fit to the differential cross-section of the form
$d\sigma/d\cost\propto 1+\cos^2\theta+
 \raisebox {0.1pc}{8}/\raisebox{-0.1pc}{3}\,\Afb\cost$,
where $\theta$ is the angle between the direction of the final 
state fermion and that of the incoming electron. This procedure
makes better use of the available information and hence 
leads to slightly smaller statistical errors. The electron final
state is special due to $t$-channel diagrams, as is discussed 
in more detail in Section\,\ref{sec-tcherr}.
The forward-backward asymmetries do not require any normalisation, 
but rely on precise measurements of the production angles of the final 
state fermions. Forward-backward asymmetries in $\qq$ final 
states are not considered here, because these require either
dedicated techniques for the tagging of quark flavours or 
a special method to extract the inclusive quark forward-backward 
asymmetry from the natural mixture of quark flavours in hadronic 
events.

In general, the systematic errors arising from the selection procedures
are small and so the accumulated statistics can be fully exploited.
Furthermore, the purely experimental errors arising from the limited
understanding of detector acceptances are uncorrelated among the experiments.
Errors arising from limitations in theoretical precision, such as the 
calculation of the 
small-angle Bhabha cross-section, the $t$-channel contribution in the $\ee$
final state or pure QED corrections to the cross-section, are common to all
experiments. They are discussed in detail in Section\,\ref{sec-comerr}.

An overview of the experimental systematic errors is given in 
Table\,\ref{tab:LSsyst}. The systematic error on the luminosity is
common to cross-section measurements of all final states, but does 
not affect the measurements of $\Afb$.

\begin{footnotesize}
\begin{table}[bht]
\begin{center} \begin{tabular}{l||ccc|ccc}
 &\mth{{\Aleph}}&\multicolumn{3}{c}{\Delphi}\\ 
\cline{2-7}
 &'93&'94&'95   &'93  &'94&'95  \\ 
\hline
\hline
${\cal L}^{\rm exp.}$
 &0.067\%&0.073\%&0.080\%&0.24\%&0.09\% &0.09\% \\
\hline
$\sigma_{\rm {had}}$
&0.069\% &0.072\%& 0.073\% &0.10\%& 0.11\% &0.10\% \\ 
$\sigma_{\rm e}$
&0.15\% &0.13\%& 0.15\%& 0.46\% &0.52\% &0.52\% \\
$\sigma_{\mu}$
&0.11\% &0.09\%& 0.11\%& 0.28\% &0.26\% &0.28\% \\
$\sigma_{\tau}$
&0.26\% &0.18\%& 0.25\%&0.60\% &0.60\%&0.60\%  \\
\hline 
\hline 
$\Afb^{\rm e}$
   &0.0006 &0.0006 &0.0006 &0.0026&0.0021 &0.0020 \\
$\Afb^{\mu}$
   &0.0005 &0.0005 &0.0005 &0.0009&0.0005 &0.0010 \\
$\Afb^{\tau}$
   &0.0009 &0.0007 &0.0009 &0.0020&0.0020& 0.0020 \\

\multicolumn{7}{c}{~} \\

&\mth{{\Lthree}}&\multicolumn{3}{c}{\Opal}\\
\cline{2-7}
&'93  &'94 &'95  &'93  &'94 &'95\\
\hline
\hline
${\cal L}^{\rm exp.}$
&0.086\%& 0.064\%&0.068\%&0.033\%&0.033\%&0.034\%\\
\hline
$\sigma_{\rm {had}}$
&0.042\%& 0.041\%& 0.042\%& 0.073\%& 0.073\%&0.085\%\\
$\sigma_{\rm e}$
&0.24\%& 0.17\%&0.28\% &0.17\%&0.14\%& 0.16\%\\
$\sigma_{\mu}$
&0.32\%&0.31\%&0.40\%&0.16\%&0.10\%& 0.12\%\\
$\sigma_{\tau}$
&0.68\%&0.65\%&0.76\%&0.49\%&0.42\%&0.48\%\\
\hline 
\hline 
$\Afb^{\rm e}$
&0.0025&0.0025   & 0.0025&0.001&0.001& 0.001 \\
$\Afb^{\mu}$
&0.0008&0.0008& 0.0015& 0.0007&0.0004& 0.0009\\
$\Afb^{\tau}$
&0.0032&0.0032& 0.0032& 0.0012&0.0012& 0.0012\\
\end{tabular}\end{center}
\caption[Experimental systematic errors]{\em \label{tab:LSsyst}
Experimental systematic errors for the analysis of the
$\Zzero$ line shape and lepton forward-backward asymmetries at the 
$\Zzero$ peak. None of the common errors discussed in 
Section\,\ref{sec-comerr} are included here.}
\end{table}
\end{footnotesize}

\vspace*{-0.5pc}
\subsection{Energy calibration \label{sec-ecal}}
Precise knowledge of the centre-of-mass energy is essential for the
determination of the mass and width of the $\Zzero$ resonance. The key
features of the energy calibration procedure were the technique of resonant
depolarisation and the careful monitoring of all relevant machine 
parameters\,\cite{bib-ECAL95}.
The latter is important because beam energy calibrations with resonant
depolarisation were possible only outside normal data taking, usually at 
the end of data taking in a particular fill of the accelerator, with 
fills typically lasting approximately 10\,h. About 40\% of the recorded 
off-peak luminosity was calibrated in this way in the 1993 scan and about 
70\% in the 1995 scan. 
For each experiment a value of the beam energy was provided every 15 minutes.
These values were evaluated from the time evolutions of the 
relevant machine parameters. 
This required a model which took into account the fields in the {\Lep} 
dipoles and in the corrector magnets, beam orbit positions, collision offsets 
at the interaction points and parameters of the radio-frequency system. In 
addition, environmental effects from leakage currents produced by 
trains in the Geneva area and the gravitational forces of the Moon and the 
Sun leading to small deformations of the accelerator geometry had to be
considered. Errors on the centre-of-mass energy are largely dominated by 
the uncertainties in this model. The energy errors vary slightly among 
the interaction points, largely due to different configurations of the 
radio frequency cavities. The energy errors for different experiments and 
periods of data taking have large common parts, and therefore the use of 
a full correlation matrix is necessary. 
Assuming that all experiments contribute with the same weight 
allows all the {\Lep} energy errors to be conveniently summarised
in a single error matrix, common to all interaction points, as given
in Table\,\ref{tab:ECAL}.

The energy of individual beam particles is usually not at the mean 
value considered above, but oscillates around the mean beam energy. Therefore 
observables are not measured at a  sharp energy, $E^0_{cm}$, but instead 
their values are averaged over a range in energies 
$E^0_{cm} \pm \delta E_{cm}$. 
With the assumption of a Gaussian shape of the energy distribution,
the total cross-sections receive a correction proportional
to $\delta E_{cm}^2$ and the second derivative of 
$\sigma(E_{cm})$ w.r.t. $E_{cm}$. 
At {\Lep}~I, typical values of the centre-of-mass energy spread are around
50~$\MeV$. The effects of the correction lead to an increase of the 
cross-section at the peak of the $\Zzero$ resonance 
by 0.16\% and a decrease of the width by about 5~$\MeV$. The 
uncertainties on the energy spread, around $\pm 1$\,$\MeV$ 
in 1993--1995, constitute 
a negligible source of error common to all experiments. 

\begin{table}[htbp]
 \begin{center} \begin{tabular} {l|ccccccc}
 ~      & `93  & `93 & `93 &`94 & `95  & `95 & `95 \\    
 ~      &  p-2 &  p  & p+2 &  p &  p-2 &  p  & p+2 \\    
\hline 
`93 p-2 & 3.42 &     &     &     &     &     &    \\
`93 p   & 2.76 &6.69 &     &     &     &     &    \\
`93 p+2 & 2.59 &2.64 &2.95 &     &     &     &    \\    
`94 p   & 2.25 &2.38 &2.16 &3.62 &     &     &    \\    
`95 p-2 & 1.29 &1.14 &1.23 &1.23 &1.78 &     &    \\    
`95 p   & 1.19 &1.20 &1.25 &1.30 &1.24 &5.39 &    \\    
`95 p+2 & 1.20 &1.15 &1.33 &1.24 &1.22 &1.34 &1.68 \\
\end{tabular} \end{center} 
\vskip -1.5 pc \caption[Energy calibration results]
{\em \label{tab:ECAL}
Signed square root of covariance matrix elements, $({\cal V}^E)$, 
in $\MeV$, from the determination of the centre-of-mass energies for 
the scan points in 1993--1995\,\cite{bib-ECAL95}. 
Elements above the diagonal are left blank for simplicity.
The errors for the earlier years may be found in 
References\,\citen{bib-MZpaper,bib-ECAL92}.}
\end{table}

In addition to the natural energy spread, changes in the mean beam
energy due to changes of machine parameters have a similar effect. 
Periods of data taking with a very similar
centre-of-mass energy were combined into a single energy point in the
experimental analyses by performing a luminosity-weighted average. The
additional energy spread resulting from this was only around 
10~$\MeV$, which adds in quadrature to the natural  
energy spread of the accelerator.      

\vspace*{-1.pc}
\section{Parametrisation of the differential cross-section \label{sec-zpar}}

The differential cross-section for fermion pair production around
the $\Zzero$ resonance consists of three s-channel contributions: 
from $\Zzero$ exchange, photon exchange and from the 
interference between photon and $\Zzero$ diagrams,
$\sigma(s)=\sigma^{\rm z} + \sigma^{\gamma} + \sigma^{\rm int}$. 
This can be cast into a Born-type structure with
complex-valued, $s$-dependent form factors describing the couplings
of the $\Zzero$ and the photon to fermions. In the Z pole approximation, 
valid for $s \simeq \MZ^2$, these are taken to be constants.
Neglecting initial and final state photon radiation, final state 
gluon radiation and fermion masses, the 
electroweak kernel cross-section can thus be written as
\begin{displaymath} \begin{array}{ll}
\lefteqn{
\frac{2s}{\pi}\frac{1}{N_c^{\rm f}}\frac{d\sigma_{\rm ew}}{d\cost}(\eeff)~=}&\\
                                                              &    \\ [-2mm]
 & \underbrace{
\left| \alpha(\MZ) Q^{\rm f} \right|^{2} (1+\cos^{2}\theta) 
                }_{\textstyle \gamma}                 \\
                                                              &    \\ [-2mm]
 & \underbrace{
 -8\, {\rm Re} \left\{ \alpha^*(\MZ) Q^{\rm f} \chi (s)  
 \left[ \cgve\cgvf (1+\cos^{2}\theta)
 + 2 \cgae\cgaf \cost  \right] \hspace*{-0.2em}\right\} 
                      }_{\textstyle \gamma-\Zzero ~ {\rm~interference}}   \\
                                                               &  \\ [-2mm]
 &  +16\,|\chi(s)|^{2}
 \left[\, (|\cgve|^2+|\cgae|^2)(|\cgvf|^2+|\cgaf|^2)(1+\cos^{2}\theta) \right.
 \\
                                                   
 & \underbrace{
 \left. \hspace*{13ex}+ 8\,{\rm Re}\left\{\cgve{\cgae}^*\right\}
                         {\rm Re}\left\{\cgvf{\cgaf}^*\right\}
                        \cos\theta \,\right] \hspace*{9ex}  
                        }_{\textstyle \Zzero} \\

\end{array}\end{displaymath}
\vskip-0.5pc \noindent with
\vskip-1pc \[
\chi(s) = \frac{\GF\MZ^{2}}{8\pi\sqrt{2}}\,
       \frac{s}{s-\MZ^{2} 
           + {\raisebox{0.2pc}{is\GZ}}/{\raisebox{-0.2pc}{\MZ}}}~. \]
Here $\alpha(\MZ)$ is the electromagnetic coupling constant at the scale of
the $\Zzero$ mass, $\GF$ is the Fermi constant, $Q^{\rm f}$ is the charge of
the final state fermion, and the colour factor $N_c^{\rm f}$ is one for 
leptons (f=e,\,$\mu$,\,$\tau$) and three for quarks (f=u,\,d,\,s,\,c,\,b).
The effective vector and axial 
vector couplings of fermions to the $\Zzero$ are denoted by
$\cgvf$ and $\cgaf$. $\chi(s)$ is the propagator 
term characterized by a Breit-Wigner denominator with 
an $s$-dependent width. 

In Bhabha final states, $\ee\rightarrow\ee$, the $t$-channel 
diagrams also contribute to the
cross-sections, with a dominant contribution at large $\cost$,
{\sl i.\,e.} in the forward direction. This contribution, as well as its
interference with the $s$-channel, add to the pure $s$-channel 
cross-section for $\ee\rightarrow\ee$.

The $1+\cos^2\theta$ terms in the above formula 
contribute to the total cross-section, whereas the terms multiplying 
$\cost$ contribute only to the forward-backward asymmetries for an
experimental acceptance symmetric in $\cost$. The total cross-section
is completely dominated by $\Zzero$ exchange. The $\gamma - \Zzero$
interference determines the energy dependence of the forward-backward
asymmetries and dominates at off-peak energies, but the leading
contribution from the real parts of the couplings vanishes at 
$\sqrt{s}=\MZ$.

The inclusion of higher-order electroweak corrections is absorbed
in $\cga$ and $\cgv$ with small imaginary parts arising 
from electroweak form factors.  The experimental measurements do not
allow a simultaneous extraction of the real and imaginary parts,
and therefore the effective couplings to be determined are defined as 
$\ga\, = \,{\rm Re}(\cga)$ and 
$\gv\, = \,{\rm Re}(\cgv)$. 
The imaginary parts of $\cga$ and $\cgv$ are explicitly accounted 
for in the fitting codes by setting them to their SM expectations. 
The effects of box diagrams are also taken into account at this level. 

It is worth noting that the definitions of the mass and width with 
an $s$-dependent width term in the Breit-Wigner denominator 
are suggested by phase-space and the structure of the 
electroweak radiative corrections within the SM. They differ 
from other commonly used definitions, {\sl e.\,g.} the real part of
the pole position in the energy-squared plane, where the 
propagator term takes the functional form 
$\chi(s)\propto s/(s- {\overline{\rm m}_{\rm Z}}^2
 + i\,\overline{\rm m}_{\rm Z} \overline{\Gamma}_{\rm Z})$\,.
This gives an identical line shape if
$\overline{\rm m}_{\rm Z}$ and $\overline{\Gamma}_{\rm Z}$ are
related to $\MZ$ and $\GZ$ by the multiplicative factor
{\raisebox{0.2pc}{$1$}}/{\raisebox{-0.2pc}{$\sqrt{1+\GZ^2/\MZ^2}$}}\,.

Photon radiation from 
the initial and final states, and their interference, is conveniently
treated by convoluting the electroweak kernel cross-section, 
$\sigma_{ew}(s)$, with a QED radiator, $H^{\rm tot}_{\rm QED}$, 
 $$\sigma(s) = 
   \int_{4{\rm m}_{\rm f}^2}^{1} dz\,
   H^{\rm tot}_{\rm QED}(z,s) \, \sigma_{ew}(zs)\,.$$
The difference between the forward and backward cross-sections 
entering into the determination of the forward-backward asymmetries,
$\sigma_{\rm F}-\sigma_{\rm B}$, is treated in the same way using a 
radiator function $H^{\rm FB}_{\rm QED}$.

At the peak the QED de-convoluted total cross-section is 
36\,\% larger than the measured one, and the peak position
is shifted downwards by about 100\,\MeV. The estimated precision
of this important correction is discussed in 
Section\,\ref{sec-QEDerr}. 

The partial Z decay widths are defined inclusively, {\sl i.\,e.} they
contain QED and QCD final state corrections and the contribution
from the imaginary parts of the effective couplings,  
$$ \Gff = N_c^{\rm f} \, \frac {\GF \, \MZ^3} {6\sqrt{2}\pi} \, 
      \left( |\cgaf|^2 \Raf \, + \, |\cgvf|^2 \Rvf \right) 
         \, + \, \Delta_{ew/QCD}\,. $$ 
The radiator factors $\Rvf$ and $\Raf$ take into account final state 
QED and QCD corrections as well as non-zero fermion 
masses; $\Delta_{ew/QCD}$  accounts for non-factorizable electroweak/QCD
corrections. The inclusion of the imaginary parts of the couplings in
the definition of the leptonic width, $\Gl$, leads to changes of 
0.15~per-mil corresponding to only 15\,\% of the  \Lep-combined 
experimental error on $\Gl$. 

The total cross-section arising from the $\cost$-symmetric
$\Zzero$ production term can also be written in terms of the partial 
decay widths into the initial and final states, $\Gee$ and $\Gff$, 
$$\sigma_{\rm ff}^{\rm z}\,=\,
\sigma_{\rm ff}^{\rm peak}\,
      \frac{s\,\GZ^2} 
{(s-\MZ^2)^2 + {\raisebox{0.2pc}{$s^2\GZ^2$} / \raisebox{-0.2pc}{$\MZ^2$}} },$$
$$~{\rm with}~ 
\sigma_{\rm ff}^{\rm peak}=\frac{1}{1+\delta_{\rm QED}}\sigma_{\rm ff}^0 
~{\rm and}~ 
\sigma_{\rm ff}^0=\frac{12\,\pi}{\MZ^2}\ \frac{\Gee\Gff}{\GZ^2}\,.$$
The term $1/(1+\delta_{\rm QED})$ removes the final state QED 
correction included in the definition of $\Gee$. 

No distinction of the flavours of produced quarks is made, and therefore
the overall hadronic cross-section is measured, and is parameterised in
terms of the hadronic width given by the sum over all quark final
states, $$\Ghad=\sum_{{\rm q,\,q}\neq{\rm t}}^{} \Gamma_{\qq}\,.$$

The decays of the $\Zzero$ to neutrinos are invisible in the detectors 
and give rise to the ``invisible width'', 
$\Ginv=N_{\nu}\Gnu$, where $N_{\nu}$ is the number 
of light neutrino species. The invisible width can be determined from 
the measurements of the decay widths to all visible final states 
and the total width, which is given by the sum over all partial widths, 
$$\GZ=\Gee+\Gmumu+\Gtautau+\Ghad+\Ginv\,.$$ 

Because the measured cross-sections depend on products of the partial 
widths and also on the total width, the widths constitute a highly 
correlated parameter set. In order to reduce correlations among 
the fit parameters an experimentally-motivated set of six parameters 
is used to describe the total hadronic and leptonic cross-sections 
around the $\Zzero$ peak. These are
\begin{itemize}
\item the mass of the $\Zzero$, $\MZ$, and the total width, $\GZ$;
\item the ``hadronic pole cross-section'',
\[ \shad\equiv{12\pi\over\MZ^2}{\Gee\Ghad\over\GZ^2} \, ; \]
\item the ratios
\[ \Ree\equiv\Ghad/\Gee\,,~\Rmu\equiv\Ghad/\Gmumu 
  ~~{\rm and}~~ \Rtau\equiv\Ghad/\Gtautau \, . \]
\end{itemize}
The leading contribution from $\gamma$-$\Zzero$ interference 
is proportional to the product of the vector couplings of the
initial and final states and vanishes at $\sqrt{s}=\MZ$, but
becomes noticeable at off-peak energies and therefore
affects the $\Zzero$ mass. Because a determination of all quark 
couplings is not possible, the $\gamma$-$\Zzero$ interference term in 
the hadronic final state is fixed to its SM value. The implications
of this are discussed in Section\,\ref{sec-jhad}.

Three additional parameters are needed to describe the leptonic
forward-backward asymmetries for the processes 
$\eeee$, $\eemumu$ and $\eetautau$. These are 
\begin{itemize}
\item the ``pole asymmetries'',  $\Afbze$, $\Afbzm$ and $\Afbzt$.
\end{itemize} 
Contrary to the partial widths, the pole asymmetries are defined 
purely in terms of the real parts of the effective $\Zzero$ couplings,
$$
\Afbzf   \equiv  {3\over 4} \cAe\cAf  ~~{\rm with}~~ 
   \cAf=\frac{2 \, {\raisebox{0.2pc}{$\gvf$}/\raisebox{-0.2pc}{$\gaf$}}}
    {1\,+\,
  \left(\raisebox{0.2pc}{$\gvf$}\, / \,\raisebox{-0.2pc}{$\gaf$}\right)^2}\,.
$$
Due to the smallness of the leptonic forward-backward asymmetry at 
$\sqrt{s}=\MZ$, QED corrections are as large as $\Afbpol$ itself. 
The product of the axial vector couplings of the initial and final 
states determines the leading contribution of the $\gamma$-$\Zzero$ 
interference. This can be fixed with sufficient precision together 
with the vector couplings from simultaneous fits to the measured 
forward-backward asymmetries and cross-sections, requiring SM input 
only for the imaginary parts of the couplings. 

Differences between the pseudo-observables
and the QED de-convoluted observables at $\sqrt{s}=\MZ$, arising
from the interference between photon 
and $\Zzero$ diagrams and from the interplay between the real and 
imaginary parts of the photon and $\Zzero$ couplings, are small in
the SM. $\sigma_{\rm ff}^{\rm peak}$, given in terms of the partial 
decay widths, agrees to better than 0.5\,per-mil for both 
hadrons and leptons with the QED de-convoluted
cross-sections without the photon exchange contribution
at $\sqrt{s}=\MZ$. This is only a small 
fraction of the {\Lep}-combined experimental error. The difference 
between $\Afbpol$ and the QED de-convoluted forward-backward asymmetry 
at the peak amounts to 0.0013, which is slightly larger than the 
{\Lep}-combined error on $\Afbpol$. It is therefore important to 
treat the imaginary parts correctly, however, the measurements are 
not sensitive to variations of the imaginary parts within their
SM expectation.

The pseudo-observables introduced above cannot 
be considered as truly model-in\-de\-pen\-dent, because imaginary parts 
of the couplings as well as the $\gamma$-$\Zzero$ interference in the 
hadronic final state need to be fixed to their SM values. 
This leads to small ``Standard Model remnants'' in any attempted 
``model-independent'' definition of the pseudo-observables. 
More details about the treatment of imaginary parts and SM remnants
in the theory programs TOPAZ0\,\cite{bib-TOPAZ0} 
and ZFITTER\,\cite{bib-ZFITTER} are given in Reference\,\citen{PCP99}.  
These computer codes make available the best current knowledge of 
QED and electroweak corrections within the minimal Standard Model and 
also provide a (quasi) model-independent approach based on the parametrisation 
of the cross-sections and forward-backward asymmetries, 
as described above.

\section{Experimental results \label{sec-inpdata}}
The experimental results presented here have been  
slightly modified from those published by the 
experiments~\cite{ALEPHLS,DELPHILS,L3LS,OPALLS}
in order to facilitate the combination procedure. The four 
dedicated sets of experimental results for the combination are 
summarised in Table\,\ref{tab:Zparinput}. 

\begin{table}[p] \begin{center}{\small
\begin {tabular} {lr|@{\,}r@{\,}r@{\,}r@{\,}r@{\,}r@{\,}r@{\,}r@{\,}r@{\,}r}
\multicolumn{2}{c|}{~}& \multicolumn{9}{c}{correlations} \\
\multicolumn{2}{c|}{~} & $\MZ$ & $\GZ$ & $\shad$ &
     $\Ree$ &$\Rmu$ & $\Rtau$ & $\Afbze$ & $\Afbzm$ & $\Afbzt$ \\
\hline 
\multicolumn{2}{l}{ $\pzz \chi^2/N_{\rm df}\,=\, 169/176$}& 
                                      \multicolumn{9}{c}{\Aleph} \\
\hline 
 $\MZ$\,[\GeV{}]\hspace*{-.5pc} & 91.1891 $\pm$ 0.0031     &   
  1.00 \\
 $\GZ$\,[\GeV]\hspace*{-2pc}  &  2.4959 $\pm$ 0.0043     & 
  .038 & ~1.00 \\ 
 $\shad$\,[nb]\hspace*{-2pc}  &  41.558 $\pm$ 0.057$\pz$ &   
 $-$.091 & $-$.383 & ~1.00 \\
 $\Ree$        &  20.690 $\pm$ 0.075$\pz$ &   
  .102  &~.004 & ~.134 & ~1.00 \\
 $\Rmu$        &  20.801 $\pm$ 0.056$\pz$ &   
 $-$.003 & ~.012 & ~.167 & ~.083 & ~1.00 \\ 
 $\Rtau$       &  20.708 $\pm$ 0.062$\pz$ &   
 $-$.003 & ~.004 & ~.152 & ~.067 & ~.093 & ~1.00 \\ 
 $\Afbze$      &  0.0184 $\pm$ 0.0034     &   
 $-$.047 & ~.000 & $-$.003 & $-$.388 & ~.000 & ~.000 & ~1.00 \\ 
 $\Afbzm$      &  0.0172 $\pm$ 0.0024     &   
 .072 & ~.002 & ~.002 & ~.019 & ~.013 & ~.000 & $-$.008 & ~1.00 \\ 
 $\Afbzt$      &  0.0170 $\pm$ 0.0028     &   
 .061 & ~.002 & ~.002 & ~.017 & ~.000 & ~.011 & $-$.007 & ~.016 & ~1.00 \\
    ~          & \multicolumn{2}{c}{~}                \\[-0.5pc]
\hline 
\multicolumn{2}{l}{$\pzz \chi^2/N_{\rm df}\,=\, 177/168$} & 
                                        \multicolumn{9}{c}{\Delphi} \\
\hline 
 $\MZ$\,[\GeV{}]\hspace*{-.5pc}   &  91.1864 $\pm$ 0.0028    &
 ~1.00 \\ 
 $\GZ$\,[\GeV]\hspace*{-2pc}   &  2.4876 $\pm$ 0.0041     &
 ~.047 & ~1.00 \\ 
 $\shad$\,[nb]\hspace*{-2pc}   &  41.578 $\pm$ 0.069$\pz$ &
 $-$.070 & $-$.270 & ~1.00 \\ 
 $\Ree$        &  20.88  $\pm$ 0.12$\pzz$ &
 ~.063 & ~.000 & ~.120 & ~1.00 \\ 
 $\Rmu$        &  20.650 $\pm$ 0.076$\pz$ &
 $-$.003 & $-$.007 & ~.191 & ~.054 & ~1.00 \\ 
 $\Rtau$       &  20.84  $\pm$ 0.13$\pzz$ &
 ~.001 & $-$.001 & ~.113 & ~.033 & ~.051 & ~1.00 \\ 
 $\Afbze$      &  0.0171 $\pm$ 0.0049     &
 ~.057 & ~.001 & $-$.006 & $-$.106 & ~.000 & $-$.001 & ~1.00 \\ 
 $\Afbzm$      &  0.0165 $\pm$ 0.0025    &
 ~.064 & ~.006 & $-$.002 & ~.025 & ~.008 & ~.000 & $-$.016 & ~1.00 \\ 
 $\Afbzt$      &  0.0241 $\pm$ 0.0037     & 
 ~.043 & ~.003 & $-$.002 & ~.015 & ~.000 & ~.012 & $-$.015 & ~.014 & ~1.00 \\
    ~          & \multicolumn{2}{c}{~}                \\[-0.5pc]
\hline 
\multicolumn{2}{l}{$\pzz \chi^2/N_{\rm df}\,=\, 158/166 $}  & 
                                    \multicolumn{9}{c}{\Lthree} \\
\hline %
 $\MZ$\,[\GeV{}]\hspace*{-.5pc}   &  91.1897 $\pm$ 0.0030      & 
 ~1.00 \\ 
 $\GZ$\,[\GeV]\hspace*{-2pc}   &   2.5025 $\pm$ 0.0041      & 
 ~.065 & ~1.00 \\ 
 $\shad$\,[nb]\hspace*{-2pc}   &   41.535 $\pm$ 0.054$\pz$  & 
 ~.009 & $-$.343 & ~1.00 \\ 
 $\Ree$        &   20.815  $\pm$ 0.089$\pz$ & 
 ~.108 & $-$.007 & ~.075 & ~1.00 \\ 
 $\Rmu$        &   20.861  $\pm$ 0.097$\pz$ & 
 $-$.001 & ~.002 & ~.077 & ~.030 & ~1.00 \\ 
 $\Rtau$       &   20.79 $\pz\pm$ 0.13$\pzz$& 
 ~.002 & ~.005 & ~.053 & ~.024 & ~.020 & ~1.00 \\ 
 $\Afbze$      &   0.0107 $\pm$ 0.0058      & 
 $-$.045 & ~.055 & $-$.006 & $-$.146 & $-$.001 & $-$.003 & ~1.00 \\ 
 $\Afbzm$      &   0.0188 $\pm$ 0.0033      & 
 ~.052 & ~.004 & ~.005 & ~.017 & ~.005 & ~.000 & ~.011 & ~1.00 \\ 
 $\Afbzt$      &   0.0260 $\pm$ 0.0047      & 
 ~.034 & ~.004 & ~.003 & ~.012 & ~.000 & ~.007 & $-$.008 & ~.006 & ~1.00 \\
    ~          & \multicolumn{2}{c}{~}                \\[-0.5pc]
\hline 
\multicolumn{2}{l}{$\pzz \chi^2/N_{\rm df}\,=\, 155/194 $} & 
                                      \multicolumn{9}{c}{\Opal} \\
\hline %
 $\MZ$\,[\GeV{}]\hspace*{-.5pc}   & 91.1858 $\pm$ 0.0030 &
 ~1.00 \\ 
 $\GZ$\,[\GeV]\hspace*{-2pc}   & 2.4948  $\pm$ 0.0041 & 
 ~.049 & ~1.00 \\ 
 $\shad$\,[nb]\hspace*{-2pc}   & 41.501 $\pm$ 0.055$\pz$ & 
 ~.031 & $-$.352 & ~1.00 \\ 
 $\Ree$        & 20.901 $\pm$ 0.084$\pz$ &
 ~.108 & ~.011 & ~.155 & ~1.00 \\ 
 $\Rmu$        & 20.811 $\pm$ 0.058$\pz$ &
 ~.001 & ~.020 & ~.222 & ~.093 & ~1.00 \\ 
 $\Rtau$       & 20.832 $\pm$ 0.091$\pz$ &
 ~.001 & ~.013 & ~.137 & ~.039 & ~.051 & ~1.00 \\ 
 $\Afbze$      & 0.0089 $\pm$ 0.0045 &
 $-$.053 & $-$.005 & ~.011 & $-$.222 & $-$.001 & ~.005 & ~1.00 \\ 
 $\Afbzm$     & 0.0159 $\pm$ 0.0023 &
 ~.077 & $-$.002 & ~.011 & ~.031 & ~.018 & ~.004 & $-$.012 & ~1.00 \\ 
 $\Afbzt$      & 0.0145 $\pm$ 0.0030 & 
 ~.059 & $-$.003 & ~.003 & ~.015 & $-$.010 & ~.007 & $-$.010 & ~.013 & ~1.00 \\
\end{tabular}}
\end{center}
\vskip-1.5pc\caption[Nine parameter results]{\label{tab:Zparinput} \em \small
Results on Z parameters and their correlation coefficients. }
\end{table} 

All fits are based on versions 6.23 of ZFITTER  and 4.4 of TOPAZ0. 
The published {\Aleph} results were derived using version 6.10 of ZFITTER,
which did not yet contain the improved treatment of fermion pairs
radiated from the initial state~\cite{Arbuzov}. 
For the combination presented here, the ALEPH measurements were    
reanalysed using version 6.23 of ZFITTER.

Each experiment used the combined energy error matrix described above 
(Table\,\ref{tab:ECAL}). This makes a small difference at the level
of 0.1\,{\MeV} on $\MZ$ and $\GZ$ and their errors only for L3,
where uncertainties arising from the modelling of the radio frequency 
cavities are largest.

In the Bhabha final state, the $s$-$t$ interference has a 
small dependence on the value of the $\Zzero$ mass. Although
practically negligible for a single experiment, a consistent 
treatment becomes important for the combination. Despite
some different choices in the publications of the individual
analyses, all experiments evaluate the $t$,\,$s$-$t$ channel 
correction at their own value of the $\Zzero$ mass for the
results presented here. The resulting interdependencies
between the $\Zzero$ mass and the parameters from the Bhabha
final state are explicitely included in the error correlation 
coefficients between $\MZ$ and $\Ree$ or $\Afbze$.

The {\Lep} experiments agreed to use a standard set of 
parameters for the calculation of the Standard Model 
remnants in the theory programs. The important parameters are 
the $\Zzero$ mass, $\MZ=91.187\,\GeV$, 
the Fermi constant, 
$\GF=(1.16637\pm0.00001)\times 10^{-5}~\GeV^{-2}$~\cite{bib-GFnew},
the electromagnetic coupling constant,
$\alpha^{(5)}(\MZ)=1/128.877\pm0.090$\,\footnote{ 
$\alpha^{(5)}(\MZ)$ is the electromagnetic coupling constant at
the scale of the $\Zzero$ mass for five quark flavours; the value
and error given correspond to a correction due to 
hadronic vacuum polarisation of 
$\Delta\alpha^{(5)}_{\rm had}=0.02804\pm0.00065$.}~\cite{bib-ALPHAMZ},
the strong coupling constant, $\alpha_s(\MZ)=0.119\pm0.002$~\cite{bib-pdg98},
the top quark mass, $\Mt=174.3\pm5.1\,\GeV$~\cite{bib-pdg2000}, 
and finally the Higgs mass, $\MH$, which was fixed to 150\,\GeV{}. 
The dependence of the fit results arising from the uncertainties in 
these parameters is almost negligible, as is discussed
in Section\,\ref{sec-parerr}. 

\vspace{\fill}

\section{Common uncertainties \label{sec-comerr}}
Important common errors among the results from all {\Lep} experiments 
arise from several sources. These are the calibration
of the beam energy, the theoretical error on the calculation
of the small-angle Bhabha cross-section used as the normalisation
reaction for all cross-section measurements, the theoretical 
uncertainties in the $t$-channel and $s$-$t$ interference contribution
to the differential large-angle Bhabha cross-section, the theoretical
uncertainties in the calculations of QED radiative effects and, finally,
from small uncertainties in the parametrisation of the electroweak 
cross-section near the $\Zzero$ resonance in terms of the set of 
pseudo-observables the four collaborations agreed upon. 
These common errors are quantified in the following sub-sections
and are used in the combination.

Other sources of common errors may arise from the use of
common Monte Carlo generators and detector simulation programs. 
However, each group uses its own tuning procedures and event
selections which best suit their detector, and therefore the
related errors are treated as uncorrelated among the experiments. 

\subsection{Common energy uncertainties \label{sec-comenerr}}
For the purpose of combining the experimental results at the parameter 
level, the errors on the 
centre-of-mass energy of each individual cross-section or asymmetry
measurement, as given in Table\,\ref{tab:ECAL}, 
need to be transformed into errors on the extracted pseudo-observables.
The first step is to scale the energy errors by factors of $1\pm\epsilon$, 
while maintaining the experimental errors fixed. Typical values of 
$\epsilon$ used are between 5\,\% and 20\,\%.
Performing the standard fits with these scaled errors
generates two pseudo-observable covariance matrices, $V_\pm$, from which
the covariance matrix due to energy errors, $V_E$, can be separated from the
other errors, $V_{\rm exp}$, using the relation
$(V_\pm) = (1 \pm \epsilon)^2 (V_E) + (V_{\rm exp})$. The validity of this
procedure was verified using a data set restricted to the hadronic 
cross-section measurements of the years 1993--1995, which were combined 
both at the cross-section level and at the parameter level.

Table\,\ref{tab:cove9ap} and \ref{tab:cove_afbap} in the 
appendix show the energy errors on the pseudo-observables 
extracted from the individual experimental data sets. 
The estimated energy errors differ slightly depending on which 
experimental data set is used to derive them. Combinations may 
be attempted based on each of them, or on the average. The central 
values and errors of each of the averaged parameters agree well to
within 5\,\% of the error on that average. It is therefore most 
appropriate to take the average of the error estimates over the 
experiments as the common energy errors, which are shown in 
Table\,\ref{tab:cove9}.

\begin{table}[hbt] \begin{center} \begin{tabular}{lr}
\begin{minipage}[t]{0.57\textwidth}
\begin{tabular} {l|rrrr}
  ~     &   $\MZ\pz$     & $\GZ\pz$      & $\shad\pz$ & $\Ree\pz$ \\
\hline 
$\MZ$\,[\GeV]\hspace*{-0.3pc} &   0.0017 & & &  \\
$\GZ$\,[\GeV]\hspace*{-2pc}   &$-$0.0006 &   0.0012 &  & \\
$\shad$\,[nb]\hspace*{-2pc}   &$-$0.0018 &$-$0.0027 &   0.011$\pz$ & \\
$\Ree$                        &   0.0017 &$-$0.0014 &   0.0073 & 0.013 \\
\end{tabular}
\end{minipage}  
      &
\begin{minipage}[t]{0.4\textwidth}
\begin{tabular} {l|rrr}
    ~    & $\Afbze\pz$ & $\Afbzm\pz$ & $\Afbzt\pz$ \\
\hline 
$\Afbze$ &   0.0004 &  &  \\
$\Afbzm$ &$-$0.0003 &   0.0003 &  \\
$\Afbzt$ &$-$0.0003 &   0.0003 & 0.0003 \\
\end{tabular}
\end{minipage}  \\
\end{tabular}\end{center}
\caption {\label{tab:cove9}\em \small 
Common energy errors for nine-parameter fits. Values are given as the 
signed square root of the covariance matrix elements in the same units
as in Table\,\ref{tab:Zparinput}; elements above 
the diagonal have been omitted for sim\-pli\-city. The anti-correlation 
between electron and muon or tau asymmetries arises from the 
different energy dependence of the electron asymmetry
due to the $t$-channel contribution.} \end{table}

\subsection{Common ${\mathbf{t}}$-channel uncertainties  \label{sec-tcherr}}
The $t$-channel and $s$-$t$ interference contributions are 
calculated in the Standard Model using the program 
ALIBABA\,\cite{bib-ALIBABA}. 
The $s$-$t$ interference contribution to the $t$-channel
correction in Bhabha final states depends on the value of
the $\Zzero$ mass. For the purpose of this combination,
all experiments parametrise the $t$ and $s$-$t$ contributions as a 
function of $\MZ$. This allows the $t$,~$s$-$t$ correction to 
follow the determination of $\MZ$ in the fits, which results in a 
correlation between $\MZ$ and $\Ree$ or $\Afbze$. The change in 
correlation coefficients introduced by explicitly taking into account
the $\MZ$ dependence of the $t$ channel in the fits is about +10\,\% 
for the $\MZ$-$\Ree$ correlation and $-$10\,\% for the 
$\MZ$-$\Afbze$ one. These correlation coefficients take 
the changes in $\Ree$ and $\Afbze$ properly into account 
when $\MZ$ takes its average value in the combination of the 
four experiments. 

The theoretical uncertainty on the $t$-channel correction is discussed 
in detail in Reference\,\citen{bib-tierror1}.
The size of the uncertainty is typically 1.1 pb for the forward cross-section
and 0.3 pb for the backward cross-section and depends slightly on the
acceptance cuts\,\cite{bib-tierror2}.
All collaborations incorporate the theory uncertainty as an additional error 
on the electron pair cross-section and asymmetry.
In order to evaluate the common error from this source,
each collaboration performed two fits, with and without the 
theory error, and the quadratic differences of the covariance
matrix elements for $\Ree$ and $\Afbze$ are taken as an estimate of the 
common error. The unknown error correlation between energy points 
below and above the peak is included in the error estimates
by adding in quadrature the observed shifts in mean values of
$\Ree$ and $\Afbze$ when varying this correlation between 
$-1$ and $+1$. The  $t$,~$s$-$t$ related errors estimated by 
individual experiments are summarised in Table\,\ref{tab:tch_systap} 
in the appendix. Since these are all very similar, the average shown 
in Table\,\ref{tab:tch_syst} is taken as the common error matrix.

\begin{table}[hbt] \begin{center}
\begin{tabular} {l|rr}
     ~    & $\Ree\pzz$ & $\Afbze\pz$ \\
\hline 
 $\Ree$   &   0.024$\pz$ &            \\
 $\Afbze$ &$-$0.0054     & 0.0014     \\
\end{tabular}
\end{center}
\caption {\label{tab:tch_syst}\em \small
 Common errors arising from the $t$-channel and $s$-$t$ interference
 contributions to the $\ee$ final states, given as the signed square root 
 of the covariance matrix elements.} 
\end{table} 

\subsection{Common luminosity uncertainties \label{sec-lumerr}}
The four collaborations use similar techniques to measure the
luminosity of their data samples by counting the number of small-angle
Bhabha scattered electrons.
The experimental details of the four measurements differ sufficiently
that no correlation is believed to exist in the experimental component
of the luminosity errors.
All four collaborations, however, use 
BHLUMI~4.04\,\cite{bib-BHLUMI4}, the best available Monte Carlo 
generator for small-angle Bhabha scattering, to calculate 
the acceptance of their luminosity counters.
Therefore significant correlations exist in the errors assigned to the 
scale of the measured cross-sections due to the uncertainty in this 
common theoretical calculation.

This uncertainty is estimated to be 0.061\,\%\,\cite{bib-BHLUMI061} without
applying a correction for the production of light fermion pairs, which is
not calculated in BHLUMI, and enters as a contribution to the
estimated error.
A recent calculation of the contribution of light pairs\,\cite{bib-SOFTPAIRS}
has allowed {\Opal} to explicitly correct for light pairs and reduce its
theoretical luminosity uncertainty to 0.054\,\%.
This is taken as common with the errors of the other three experiments, who 
between themselves share a mutual common error of 0.061\,\%.

These errors almost exclusively affect the hadronic pole cross-section,
and contribute about half its total error after combination.
The common luminosity error also introduces a small contribution to 
the covariance matrix element between $\GZ$ and $\shad$.  This was 
neglected in the common error tables given above, as it had 
no noticeable effect on the combined result.

\vspace*{-0.5pc}
\subsection{Common theory uncertainties  \label{sec-therr}}
An additional class of common theoretical errors arises from the 
approximations and special choices made in the fitting codes. 
These comprise contributions from QED radiative corrections, including 
initial-state pair radiation, and the parametrisation of the differential
cross-section around the $\Zzero$ resonance in terms of pseudo-observables
defined precisely at the peak and for pure $\Zzero$ exchange only.
In order to estimate the uncertainties from the parametrisation of
the electroweak cross-sections near the $\Zzero$ resonance the two
most advanced calculational tools, ZFITTER\,\cite{bib-ZFITTER} and
TOPAZ0\,\cite{bib-TOPAZ0}, were compared. In addition, there are
``parametric uncertainties'' arising from parameters of the SM
that are needed to fix the  SM remnants.

\vspace*{-0.5pc}
\subsubsection{QED uncertainties  \label{sec-QEDerr}}
The effects of initial state photon and fermion pair radiation 
lead to the large corrections 
in the vicinity of the $\Zzero$ resonance illustrated in 
Figure\,\ref{fig:xsh_all}, and therefore play a central role 
in the extraction of the pseudo-observables from the 
measured cross-sections and asymmetries. Such large radiative 
effects have to be seen in contrast to the experimental 
precision, which is below the per-mil level in the case of the 
hadronic cross-section. 

The most up-to-date evaluations of photonic corrections include the leading
contributions up to ${\cal O}(\alpha^3)$. Two different schemes are 
available to estimate the remaining uncertainties:
\begin{enumerate}
{\parsep=0pt \itemsep=0pt \topsep=0pt \parskip=0pt \partopsep=0pt}
\item KF: ${\cal O}(... \alpha^2L^2, \alpha^2L, \alpha^2L^0)$ 
calculations\,\cite{Berends} including the exponentiation scheme of 
Kuraev-Fadin\,\cite{KF} with ${\cal O}(\alpha^3L^3)$\,\cite{Montagna}.
\footnote{Third-order terms for the KF scheme had also been 
calculated earlier\,\cite{Skrzypek}.}
\item YFS: the 2$^{\rm nd}$ order inclusive exponentiation scheme of 
Reference\,\citen{Skrzypek,JSW}, based on the YFS approach\,\cite{YFS}. Third 
order terms are known and have only a small effect\,\cite{JPS}.
\end{enumerate}

Differences between these schemes, which are both implemented in 
ZFITTER, TOPAZ0 and MIZA\,\cite{bib-MIZA}, and uncertainties due 
to missing higher order 
corrections\,\cite{JPS}, amount to at most $\pm 0.1\,\MeV$ on $\MZ$ 
and $\GZ$, and $\pm 0.01$\,\% on $\shad$.  

The influence of the interference between initial and final state radiation
on the extracted parameters has also been studied 
recently\,\cite{bib-ifi}, and uncertainties on $\MZ$ of at most 
$\pm$0.1\MeV{} from this source are expected for the experimental 
results given with only a small cut on $s'$, the effective squared 
centre-of-mass energy after photon radiation from the initial state. 
The uncertainties due to the extrapolation of the leptonic 
$s$-channel cross-sections to full angular acceptance and from 
large to small $s'$ are different among the experiments and 
are believed to be largely uncorrelated.

QED related uncertainties are dominated by the radiation of fermion 
pairs from the initial state. Starting from the full second order pair 
radiator\,\cite{Berends,ISPP88}, a simultaneous exponentiation scheme 
for radiated photons and pairs was proposed in Reference\,\citen{pairs}.
A third-order pair radiator was calculated recently\,\cite{Arbuzov}
and compared with the other existing schemes, which are all 
available in the latest version of ZFITTER. Independent implementations 
of some schemes exist in TOPAZ0 and in MIZA. The largest uncertainty 
arises from the contribution of hadronic pairs. The maximum differences
are $0.3\,\MeV$ on $\MZ$, $0.2\,\MeV$ on $\GZ$ and $0.015$\,\% on 
$\shad$. 
 
In summary, comparing the different options for photonic and fermion 
pair radiation leads to error estimates of  $\pm 0.3\,\MeV$ on $\MZ$ 
and $\pm 0.2\,\MeV$ on $\GZ$. The observed differences in $\shad$ are 
slightly smaller than the error estimate of $\pm 0.02\,$\% in 
Reference\,\citen{JPS}, which is therefore taken as the error 
for QED uncertainties.

\subsubsection{Parametrisation of line shape and asymmetries}
In a recent detailed comparison of TOPAZ0 and ZFITTER\,\cite{PCP99}, 
cross-sections from Standard Model 
calculations and from the model-independent parametrisation 
were considered. Uncertainties on the pseudo-observables arise 
from differing choices in the parametrisation of the electroweak 
cross-sections near the  $\Zzero$ resonance. In order to determine 
these TOPAZ0-ZFITTER differences, each of the two
codes have been used. For practical reasons, cross-sections and 
forward-backward asymmetries were calculated with TOPAZ0 and then 
fitted with ZFITTER. Errors were 
assigned to the calculated cross-sections and forward-backward 
asymmetries which reflected the integrated luminosity taken 
at each energy, thus ensuring that each energy point entered with 
the appropriate weight.

The dominant part of the small differences between the two codes 
results from details of the implementation of the cross-section 
parametrisation in terms of the pseudo-observables. This is 
particularly visible for the off-peak points, where the assignment
of higher-order corrections to the $\Zzero$ resonance or to the SM 
remnants is not in all cases unambiguous. The size of the differences 
also depends on the particular values of the pseudo-observables, since 
these do not neccessarily respect the exact SM relations. 
Slightly different choices are made in the two codes if the 
SM relations between the pseudo-observables are not fulfilled.  
Finally, variations of factorisation schemes and other 
options in the electroweak calculations may affect the fit results 
through the SM remnants, but were found to have a negligible effect. 

In Table\,\ref{tab:therr} differences between TOPAZ0 and ZFITTER
are shown, which are taken as common systematic errors.
They were evaluated around  the set of pseudo-observables 
representing the average of the four experiments, where the 
cross-sections and asymmetries were calculated
for full acceptance with only a cut on $s'\,>\,0.01\,s$. 
The only important systematic error of this kind is the one 
on $\Rl$, which amounts to 15\,\% of the combined error.

\begin{table}[btp] \begin{center}
\begin{tabular}{c|c|c|c|c}
$\Delta\MZ$  &$\Delta\GZ$  &$\Delta\shad$ &$\Delta\Rl$ &$\Delta\Afbpol$ \\
{}[\GeV{}]   & [\GeV]   & [nb]     &   ~      &   ~       \\
\hline 
 0.0001  &  0.0001  & 0.001    &  0.004   &  0.0001 \\
\end{tabular} \end{center} 
\caption[]{\label{tab:therr}\em \small 
Differences in fit results obtained with TOPAZ0 and ZFITTER, taken
as common systematic errors.} 
\end{table} 

Putting all sources together, overall theoretical errors as listed in
Table\,\ref{tab:QEDtherr} are obtained, and these are used in the 
combination.

\begin{table}[hbt] \begin{center}
\begin{tabular} {l|rrrrrrrrr}
  ~ &$\MZ$&$\GZ$&$\shad$&$\Ree$&$\Rmu$&$\Rtau$&$\Afbze$&$\Afbzm$&$\Afbzt$\\
\hline 
$\MZ$[\GeV{}]&0.0003& & & & & & & & \\
$\GZ$[\GeV]& ~   & 0.0002 & & & &  & \\
$\shad$[nb] & ~   &   ~   & 0.008 & & & & & &  \\
$\Ree$  & ~   &   ~   &    ~  & 0.004 & & & & & \\
$\Rmu$  & ~   &   ~   &    ~  & 0.004 & 0.004 & & & & \\     
$\Rtau$ & ~   &   ~   &    ~  & 0.004 & 0.004 & 0.004 & & & \\   
$\Afbze$ & ~  &   ~   &    ~  &  ~    &       &  ~ &0.0001& & \\
$\Afbzm$ & ~  &   ~   &    ~  &  ~    &       &  ~ &0.0001&0.0001& \\
$\Afbzt$ & ~  &   ~   &    ~  &  ~    &       &  ~ &0.0001&0.0001&0.0001\\
\end{tabular} \end{center} 
\caption[]{\label{tab:QEDtherr}\em \small 
Theoretical errors from fit programs, {\sl i.\,e.}
photon and fermion-pair radiation and model-independent
parametrisation, given as the signed square root of the 
covariance matrix elements.} 
\end{table} 

\subsubsection{Parametric uncertainties \label{sec-parerr}}
Through the SM remnants the fit results depend slightly 
on the values of some SM parameters. Varying these
within their present experimental errors,
or between 100\,\GeV{} and 1000\,\GeV{} in case of the Higgs boson 
mass, leads to observable effects only on the $\Zzero$ mass, which is 
affected through the $\gamma$-$\Zzero$ interference term. The 
dominant dependence is on $\MH$, followed by $\alpha^{(5)}(\MZ)$.

The effect on $\MZ$ from a variation of $1/\alpha^{(5)}(\MZ)$ by its 
error of $\pm$0.090 is $\mp$0.05~$\,\MeV$, which 
is negligibly small compared to the systematic error on $\MZ$ 
arising from other QED-related uncertainties 
(see Table\,\ref{tab:QEDtherr}). The change in $\MZ$ due to $\MH$ amounts to 
$+0.23\,\MeV$ per unit change in $\log_{10}(\MH/\GeV)$. Note that
this is small compared to the total error on $\MZ$ of $\pm$2.1\,{\MeV} 
and is not considered as an error, but rather as a correction to 
be applied once the Higgs boson mass is known. The consequences 
of a completely model-independent treatment of the  
$\gamma$-$\Zzero$ interference in the hadronic channel
are discussed in Section\,\ref{sec-jhad}.

\section{Combination of results \label{sec-combine}}
The combination of results on the $\Zzero$ parameters is based on 
the four sets of results on the 
nine parameters $\MZ$, $\GZ$, $\shad$, $\Ree$, $\Rmu$, $\Rtau$, 
$\Afbze$, $\Afbzm$ and $\Afbzt$ and the common errors given in the
previous chapter. 

For this purpose it is necessary to construct the 
full $(4 \times 9 )\, \times\,(4 \times 9)$ covariance matrix of the
errors. The four diagonal $9 \times 9$ matrices consist of the
four error matrices specified by each experiment 
(Table\,\ref{tab:Zparinput}). The $9 \times 9$ common error matrices 
build the off-diagonal elements. Some theoretical uncertainties must 
also be added to the diagonal matrices, since they are not contained 
in the individual experimental matrices.

A symbolic representation
of this matrix is shown in Table\,\ref{tab:covlsafb}\,. Each table 
element represents a $9 \times 9$ matrix; $({\cal C}_{exp})$ for 
$exp$~=~A, D, L and O are the covariance matrices of the 
experiments (Table\,\ref{tab:Zparinput}), and 
$({\cal C}_c) =   ({\cal C}_E) + ({\cal C}_{\cal L})
                + ({\cal C}_t) + ({\cal C}_{\rm QED,th})$ 
is the matrix of common errors. Here
$(C_E)$ (Table\,\ref{tab:cove9})
is the error matrix due to {\Lep} 
energy uncertainties, 
$(C_{\cal L})$ (Section\,\ref{sec-lumerr}) arises from the
theoretical error on the small-angle Bhabha cross-section calculations, 
$(C_t)$ (Table\,\ref{tab:tch_syst}) contains the errors 
from the $t$-channel treatment in the $\ee$
final state, and $({\cal C}_{\rm QED,th})$ 
contains the errors from initial state photon and fermion pair 
radiation and from ambiguities in the model-independent 
parametrisation (Table\,\ref{tab:QEDtherr}). 
Since the latter errors were not included in the
experimental error matrices, they were also added to the block 
matrices in the diagonal of Table\,\ref{tab:covlsafb}\,. 

\begin{table} [htb] \begin{center} \begin{tabular}{l|cccc}
 $({\cal C})$ &  {\Aleph}        & {\Delphi}        &  {\Lthree}          &  {\Opal}       \\ 
\hline
A &$({\cal C}_A)+({\cal C}_{\rm QED,th})$ & & & \\
D &$(C_c)$  & $({\cal C}_D)+({\cal C}_{\rm QED,th})$ & & \\
L &$(C_c)$  &$(C_c)$   & $({\cal C}_L)+({\cal C}_{\rm QED,th})$ & \\ 
O &$(C_c)$  &$(C_c)$   & $(C_c)$   &$({\cal C}_O)+({\cal C}_{\rm QED,th})$ \\ 
\end{tabular}\end{center}
\caption[Covariance matrix of combined line shape and asymmetry measurements]
{\em \label{tab:covlsafb} Symbolic representation of the covariance matrix,
$({\cal C})$, used to combine the line shape and asymmetry results of the 
four experiments. Elements above the diagonal are the same as those below 
and are left blank for simplicity. The components of the matrix are explained 
in the text.}
\end{table}

The combined parameter set 
and its covariance matrix are obtained from a $\chi^2$ minimisation, with 
\[
 \chi^2 = ({\bf X} - {\bf X_m})^T ({\cal C})^{-1} ({\bf X}-{\bf X_m}),
\]
where $({\bf X} - {\bf X_m})$ is the vector of residuals of the combined 
parameter set to the individual results.

Some checks of the combination procedure outlined above are described in
the following subsections, and the combined results are given in the 
tables of Section\,\ref{sec-combres}.

\subsection{Multiple-${\mathbf{\MZ}}$ fits\label{sec-elevenpar}}

In 1993 and 1995, the two years when {\Lep} performed precision
scans to measure the $\Zzero$ line shape, the experimental 
errors are very comparable, but the {\Lep} energy was appreciably 
better understood in 1995 than in 1993. In determining the 
optimum value of $\MZ$, therefore, the four experiments combined 
should give more weight to the 1995 data than they each do in 
their independent determinations. To quantify this issue 
the measurements of each experiment were fitted to determine 
independent values of $\MZ$ for the periods 1990--1992, 
1993--94 and 1995.  In this ``eleven-parameter fit'',  
each of the three mass values $\MZA$, $\MZB$ and $\MZC$  
has its specific energy error reflecting the 
different systematic errors on the absolute energy scale 
of {\Lep}. In the combination, the relative importance of energy related 
and independent experimental errors on the mass values is properly 
treated. The input and the common energy errors, estimated
in the same way as for the nine parameters, are shown in 
Section\,\ref{sec-Zpar11input}.

When the three values of $\MZ$ are condensed 
into a single one, the effects of the time dependence of 
the precision in the energy calibration is taken into account. 
The difference of $-$0.2~$\MeV$ w.r.t. the $\MZ$ value from the 
nine-parameter fits corresponds to 10\,\% of the combined error. 
All other parameters are identical to their values 
from the nine-parameter fit to within less than 5\,\% of 
the combined error. This result justifies using the 
standard combination based on the nine parameters.
Tables\,\ref{tab:cove9ap} and \ref{tab:cove_afbap} in the 
appendix show the energy errors on the pseudo-observables 
estimated from each individual experimental data set. 
The estimated energy errors differ slightly depending on which 
experimental data set is used to derive them. Combinations may 
be attempted based on each of them, or on the average. The central 
values and errors of each of the averaged parameters agree well 
within 5\,\% of the error on that average. It is therefore most 
appropriate to take the average of the error estimates over the 
experiments as the common energy errors, which are shown in 
Table\,\ref{tab:cove9}.

\begin{figure}[htb]\begin{center}
\mbox{\epsfig{file=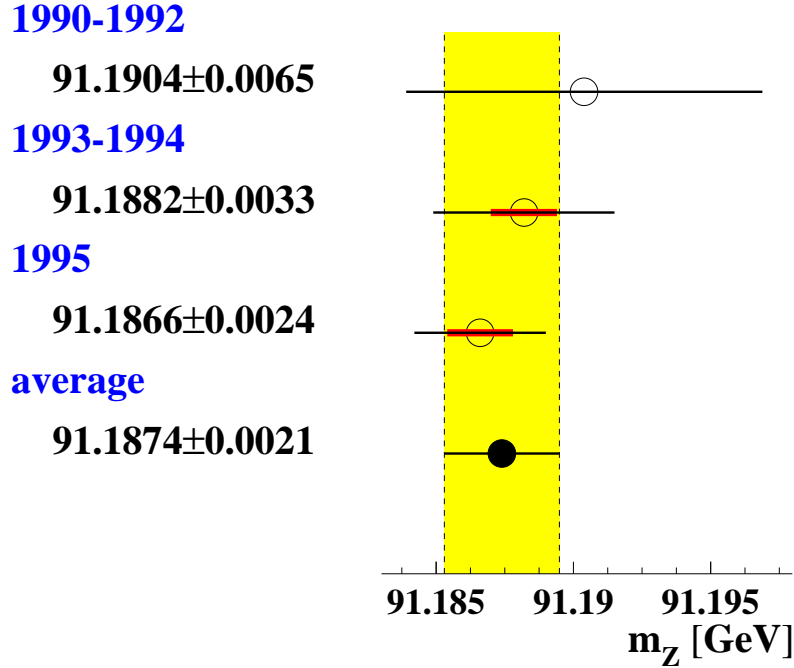,width=0.9\columnwidth}}
\end{center}
\caption[Multiple $\MZ$ fits]
{\em \label{fig:MZcheck} $\MZ$ in \GeV{} for different periods of data 
taking, before 1993, 1993--1994 and 1995.  The second, smaller
error bar represents the correlated error component of 1.2\,\MeV{}
between $\MZB$ and $\MZC$. $\MZA$ is essentially uncorrelated with
the other two.}
\end{figure}

The averages over the four experiments of the three values  
$\MZA$, $\MZB$ and $\MZC$ also provide a cross-check on the consistency of the
energy calibration, which dominates the errors on $\MZ$ in each 
of the periods considered. Using the energy errors of 
Table\,\ref{tab:cove11} allows the correlated and uncorrelated parts
of the errors on the mass differences to be quantified. This is shown in 
Figure\,\ref{fig:MZcheck}.
The differences between these values for the $\Zzero$ mass amount to
 $ | \MZA - \MZB | = 31\,\%$,   
 $ | \MZA - \MZC | = 56\,\%$ and 
 $ | \MZB - \MZC | = 43\,\%$       
 of the uncorrelated error, {\sl i.\,e.} the three $\Zzero$ mass values 
are well consistent.

\subsection{Fits with lepton universality \label{sec-fivepar}}
All experiments provide fits to their measured cross-sections and asymmetries
with lepton universality imposed, {\sl i.\,e.} $\Ree$, $\Rmu$ and $\Rtau$ 
are replaced by $\Rl$ and $\Afbze$, $\Afbzm$ and $\Afbzt$ get replaced
by $\Afbpol$ in the model-independent parametrisation of the differential
cross-section. Here $\Rl$ is defined for massless leptons.
The individual experimental results and the correlation matrices are 
given in Table\,\ref{tab:Zpar5input} in the appendix. 

Comparing these five-parameter results with the nine-parameter results
of Table\,\ref{tab:Zparinput}, there is a noticeable change in $\MZ$
of a few tenth of $\MeV$ in all experiments. This is a consequence 
of the dependence of the $t$-channel correction on $\MZ$, as discussed 
in Section\,\ref{sec-tcherr}. When $\Ree$ and $\Afbze$ are replaced by the
leptonic quantities $\Rl$ and $\Afbpol$, their correlation with the
$\Zzero$ mass leads to a shift, which is driven by the (statistical) 
difference between $\Ree$ and $\Rl$ and $\Afbze$ and $\Afbpol$. 
Similarly, replacing $\Ree$ and $\Afbze$ from the values of a single 
experiment by the {\Lep} average introduces a shift in $\MZ$ in the presence 
of these particular correlation coefficients. Such shifts should become 
smaller when averaged over the four experiments. Indeed, the average 
of the shifts is only $-0.2\,\MeV$.  

Another aspect of the five-parameter results concerns the role of
the common $t$-channel uncertainty in the averages over the leptonic
measurements to obtain $\Rl$ and $\Afbpol$.
If the average over the leptonic measurements is performed 
by each experiment individually, the weight given to the 
electron channel is larger than for the case where the averages 
over individual lepton species are averaged at the end. In the 
latter procedure the weight of the electrons relative to the muon 
and tau final states is reduced due to the common $t$-channel error.
Extracting the results with lepton universality from the nine 
parameters is therefore the appropriate method. 

\subsection{Shifts for halved experimental errors \label{sec-halvederr}}
When the average over the experiments is performed at the parameter level, 
information on the individual contribution of particular data points to
the average is lost. Performing the average over the data points instead
may therefore lead to changes of the relative importance of independent 
experimental errors w.r.t. the common errors.
The examples of $\MZ$ and the importance of the $t$-channel errors 
for $\Rl$, as discussed in the previous subsections, provide good 
illustrations of such effects. It was demonstrated that averaging the 
shifts in $\MZ$  which each experiment observed when halving its 
experimental errors to  simulate the generic  ``combined'' experiment 
also reproduced the results of the full fit to the combined hadronic 
cross-section measurements. 

While $\MZ$ is properly treated by the eleven-parameter fits,
other pseudo-observables may suffer from similar changes due to
shifts of the weights. Changes in central values when halving the 
independent experimental errors in each experiment can be used as 
a monitoring tool for these parameters as well. The average of 
these changes over the four {\Lep} experiments serves to control the 
differences between an average at the parameter level compared to a 
full cross-section average. Of course, this assumes that all 
measurements from individual experiments enter into the average 
with the same weight. The observed shifts are summarised in 
Table\,\ref{tab:halvederr}. The shift downwards in $\MZ$ of 
$0.3\,\MeV$ corresponds to the slightly smaller shift of 
$0.2\,\MeV$ already seen in the multiple-$\MZ$ fits. 

Thus, the average changes in $\MZ$, $\shad$, 
$\Ree$, $\Afbzm$ and $\Afbzt$ amount to about 10\,\% of the combined
errors, in all other cases they are even smaller.
 
\begin{table}[hbt] \begin{center} \begin{tabular}{r||rrrr|rr}
       &    A$\pzz$   &  D$\pzz$ &   L$\pzz$ &  O$\pzz$  & average & \% of error\\
\hline 
$\MZ$\,[\GeV]  &$-$0.0006  &   0.0000  &$-$0.0004  &$-$0.0001  &$-$0.00028  &13\\
$\GZ$\,[\GeV]  &$-$0.0002  &$+$0.0001  &$-$0.0004  &   0.0000  &$-$0.00013  & 5\\
$\shad$\,[nb]  &$+$0.006\pz&   0.000\pz&$+$0.008\pz&$+$0.0036  &$+$0.0037\pz&10\\
$\Ree$ &$+$0.004\pz&$+$0.017\pz&   0.000\pz&$+$0.004\pz&$+$0.0063\pz&13\\
$\Rmu$ &   0.000\pz&   0.000\pz&   0.000\pz&$+$0.001\pz&  ~0.0000\pz& 0\\
$\Rtau$&   0.000\pz&   0.000\pz&$-$0.001\pz&$+$0.002\pz&$+$0.0003\pz& 1\\
$\Afbze$&$-$0.0001 &$-$0.0003  &   0.0000  &$-$0.0000  &$-$0.00011  & 5\\
$\Afbzm$&$+$0.0002 &$+$0.0003  &   0.0000  &$+$0.0001  &$+$0.00014  &11\\
$\Afbzt$&$+$0.0002 &$+$0.0003  &   0.0000  &$+$0.0001  &$+$0.00015  & 9\\
\end{tabular}\end{center}
\caption{\label{tab:halvederr}\em \small 
 Shifts in central values of the fitted pseudo-observables seen 
when halving the independent experimental errors, for individual
experiments and average. }
\end{table}

This cross-check provides an estimate of the magnitude of 
the changes in the final results that would arise from a combination
at the cross-section level. The averaging at the parameter level is
equal to this within at most 10\,\% of the combined error.

\subsection{Influence of the ${\mathbf{\gamma}}$-${\mathbf{\Zzero}}$ 
interference term  \label{sec-jhad}}
In the nine-parameter analyses discussed here, the $\gamma$-$\Zzero$ 
interference terms in the differential cross-sections for leptons are 
expressed using the effective coupling constants and the electric charges 
of the electron, $Q^{\rm e}$, and the final state fermion, $Q^{\rm f}$ 
(see equation for the differential cross-section in Section\,\ref{sec-zpar}). 
For the hadronic final state, however, the $\gamma$-$\Zzero$ 
interference terms are fixed to the SM values, as individual 
quark flavours are not separated.  Fits with a free interference
term are possible in the S-Matrix scheme~\cite{bib-smat}.
The OPAL collaboration also studied a different approach~\cite{OPALLS}  
based on an extension of the standard parameter set.
In the S-Matrix approach the interference terms are 
considered as free and independent parameters. The hadronic interference 
term is described by the parameter $j_{\rm tot}^{\rm had}$, 
given in the SM by
\[j_{\rm tot}^{\rm had}  =
 \frac{\GF \MZ^2}{\sqrt{2}\pi\alpha(\MZ)}\, 
Q^{\rm e}\,\gve 
 \times 3\sum_{\rm q} Q^{\rm q}\,\gvq \,. \]
Note that the running of $\alpha$
as well as final state QED and QCD corrections are also included in the
definition of the S-Matrix parameters. The SM value of 
$j_{\rm tot}^{\rm had}$ is $0.21\pm0.01$.

The dependence of the nine parameters on the hadronic 
$\gamma$-$\Zzero$ interference term is studied here by
considering a set of ten parameters consisting of the standard 
nine parameters extended by the parameter $j_{\rm tot}^{\rm had}$
from the S-Matrix approach. The 
$\gamma$-$\Zzero$ interference terms in the lepton channels 
are fixed by the leptonic $\Zzero$ couplings. 
As already observed in S-Matrix analyses of the \Lep\,I 
data~\cite{L3LS,OPALLS},
a large anti-correlation between $\MZ$ and $j_{\rm tot}^{\rm had}$ 
appears, leading to errors on $\MZ$ enlarged by a factor of almost 
three. The studies show that the dependence of 
$\MZ$ on $j_{\rm tot}^{\rm had}$ is given by
\[\frac{{\rm d}\MZ}{{\rm d}j_{\rm tot}^{\rm had}}\,=\,-1.6\,\MeV/0.1\,.\]
The changes in all other parameters are below 20\,\% of the
errors in the combination for a change in 
$j_{\rm tot}^{\rm had}$ of 0.1\,.

Better experimental constraints on the hadronic interference term
are obtained by including measurements of the hadronic total 
cross-section at centre-of-mass energies further away from the Z pole
than just the off-peak energies at \Lep\,I. 
Including the measurements of the TRISTAN collaborations
TOPAZ~\cite{bib-TOPAZ} and VENUS~\cite{bib-VENUS} at $\sqrt(s)$=58\,\GeV, 
the error on  $j_{\rm tot}^{\rm had}$ is reduced to $\pm0.1$, 
while its central value is in good agreement with the SM expectation.  
Measurements at centre-of-mass energies
above the Z resonance at \Lep\,II also provide constraints on
$j_{\rm tot}^{\rm had}$, but in addition test modifications to 
the interference terms arising from the possible existence of a heavy Z' 
boson~\cite{bib-LEP2smata,bib-LEP2smatd,bib-LEP2smatl,bib-LEP2smato}.

The available experimental constraints on $j_{\rm tot}^{\rm had}$ 
thus lead to uncertainties on $\MZ$, independent of SM assumptions 
in the hadronic channel, which are already smaller than its error.
No additional error is assigned to the standard 
nine-parameter results from effects which might arise from a non-SM 
behaviour of the $\gamma$-$\Zzero$ interference.

\subsection{ Direct Standard Model fits to the measured 
cross-sections and forward-backwards asymmetries \label{sec-sm}}
 Since an important use of the combined results presented here is to
determine parameters of the SM and to make tests on its validity, it 
is crucial to verify that the parameter set we have chosen for the 
combination represents the four sets of experimental measurements 
from which they were extracted in a manner adequate to this purpose.
When the set of pseudo-observables is used in the framework of the
Standard Model, the role of $\MZ$ changes from an independent
parameter to that of a Lagrangian parameter of the theory.
This imposes additional constraints which can be expected 
to shift the value of $\MZ$.

To check whether the nine parameters adequately describe the 
reaction to these constraints, each  collaboration provided 
results from direct Standard Model fits to 
their cross-section and asymmetry data. 
The comparison of these results with those obtained by using 
the set of pseudo-observables as fit input is shown in 
Table\,\ref{tab:MZsm}. $\MH$ and $\alpha_s$ were free parameters 
in these fits, while the additional inputs $\Mt$=174.3$\pm$5.1\,\GeV{} 
and $\Delta\alpha^{(5)}_{\rm had}=0.02804\pm0.00065$
(corresponding to $1/\alpha^{(5)}(\MZ)=128.877\pm0.090$) provided 
external constraints.

Most noticeably, significant shifts in $\MZ$ are observed
in some individual data sets, which cancel to almost zero 
in the average over the four experiments. One anticipated 
source of these shifts was already mentioned: the $\Zzero$
couplings defining the $\gamma$-$\Zzero$ interference term 
depend on $\MH$, which is allowed to move freely in 
the fit for SM parameters, but is fixed to 150\,\GeV{} in the 
model-independent fit for the extraction of the pseudo-observables. 
The approximate values of 
$\MH$ preferred by the SM fit to the cross-sections and asymmetries 
are indicated in the second part of the table. Using  the dependence
of $\MZ$ on the value of $\MH$ given in Section\,\ref{sec-parerr},
the differences in $\MZ$ can be corrected to a common value of
the Higgs mass of $\MH$=150\,\GeV{}, as is shown in the last line
of Table\,\ref{tab:MZsm}. Hence the net average difference in $\MZ$ 
directly from the data or through the intermediary of the
nine parameters is less than $0.1\,\MeV$. Shifts in the 
other SM parameters, in the individual data sets as well as in the 
average, are all well under 5\,\% of the errors, and therefore also 
negligible.

\begin{table}[htb]
\begin{center}
\begin{tabular}{l||rrrr|rr}
                    &  A~~  &  D~~ &   L~~  &   O~~  & average & \% of error \\
\hline 
$\pzz \chi^2/N_{\rm df}$ & $174/180$ & 184/172 & $168/170$ & $161/198$ & ~ \\
\hline 
$\Delta\MZ$ [\MeV{}]  &$-$0.7   &$+$0.5   &0.0   &$+$0.1   &$-$0.03 & 1  \\
$\Delta\Mt$ [\GeV{}]  &   0.0   & 0.0     & 0.0  &   0.0   &   0.0  & $<$2 \\
$\Delta\log_{10}$($\MH$/\GeV{})  
                 &$-$0.01  &$+$0.04  &$+$0.02  &$+$0.04   &$+$0.02  &  4 \\
$\Delta\alpha_s$ &   0.0000&$-$0.0002&$+$0.0002&$+$0.0002 &$+$0.0001 & 4 \\
$\Delta(\Delta\alpha^{(5)}_{\rm had})$
                &$+$0.00002&$-$0.00004&   0.00000&$-$0.00004 &$-$0.00002 & 2 \\
\hline 
  fit value           &        &        &        &         &     &    \\ 
  of $\MH$ [\GeV{}]   &   40.  &   10.  &    35. &    390. &     &    \\  
\hline 
$\Delta\MZ$ [\MeV{}] &        &        &        &         &      &   \\
corrected to             &        &        &        &        &      &   \\
 $\MH$=150\,\GeV{}   &$-$0.6  &$+$0.7  &$+$0.1  &  0.0   &$+$0.05 & 2 \\
\end{tabular} \end{center}
\caption {\label{tab:MZsm}\em \small Shifts in SM parameters, from
direct SM fit to the cross-sections and forward-backward asymmetries
w.r.t. fits to the nine-parameter results.
The numbers in the lowest part of the table give the shifts in $\MZ$
if the results from the first line are corrected to a common 
value of the Higgs mass of 150\,\GeV{}.}
\end{table}

The conclusion of this study is that SM parameters extracted from
the pseudo-ob\-ser\-vab\-les are almost identical to the ones that would be
extracted from the combined cross-sections and asymmetries. Within 
the SM the combined set of pseudo-observables provides a description 
of the measurements of the $\Zzero$ parameters that is equivalent
to the full set of cross-sections and asymmetries. This is also true
for any theory beyond the SM which leads to corrections that are 
absorbed in the pseudo-observables. An exception to this are those theories
with an additional $\Zzero'$-bosons which have significant modifications
of the $\gamma$-$\Zzero$ interference term. 
(See the discussion in Section\,\ref{sec-jhad}.)

\subsection{Combined results \label{sec-combres}}

The result of the combination of the four sets of nine pseudo-observables 
of Table\,\ref{tab:Zparinput},
including the experimental and common error matrices shown in 
Table\,\ref{tab:covlsafb}, is given in Table\,\ref{tab:lsafbresult}. The 
value of the Higgs boson mass was assumed to be 150\,\GeV{} and is
relevant only for the value of $\MZ$, which changes by $+$0.23\,\MeV{} 
per unit change in $\log_{10}$($\MH$/\GeV{}). 
(See Section\,\ref{sec-parerr}.) 

\begin{table}[htb]\begin{center}
{\small 
\begin {tabular} {lr|r@{\,}r@{\,}r@{\,}r@{\,}r@{\,}r@{\,}r@{\,}r@{\,}r}
\multicolumn{2}{c|} {without lepton universality} & 
                                    \multicolumn{9}{l}{~~~correlations} \\
\hline 
\multicolumn{2}{c|}{$\pzz \chi^2/N_{\rm df}\,=\,32.6/27 $} &
   $\MZ$ & $\GZ$ & $\shad$ &
     $\Ree$ &$\Rmu$ & $\Rtau$ & $\Afbze$ & $\Afbzm$ & $\Afbzt$ \\
\hline 
 $\MZ$ [\GeV{}]  & 91.1876$\pm$ 0.0021 &
 ~1.00 \\
 $\GZ$ [\GeV]  & 2.4952 $\pm$ 0.0023 &
 $-$.024 & ~1.00 \\ 
 $\shad$ [nb]  & 41.541 $\pm$ 0.037$\pz$ &
 $-$.044 & $-$.297 & ~1.00 \\ 
 $\Ree$        & 20.804 $\pm$ 0.050$\pz$ &
 ~.078 & $-$.011 & ~.105 & ~1.00 \\ 
 $\Rmu$        & 20.785 $\pm$ 0.033$\pz$ & 
 ~.000 & ~.008 & ~.131 & ~.069 & ~1.00 \\ 
 $\Rtau$       & 20.764 $\pm$ 0.045$\pz$ &  
 ~.002 & ~.006 & ~.092 & ~.046 & ~.069 & ~1.00 \\ 
 $\Afbze$      & 0.0145 $\pm$ 0.0025 &
 $-$.014 & ~.007 & ~.001 & $-$.371 & ~.001 & ~.003 & ~1.00 \\ 
 $\Afbzm$      & 0.0169 $\pm$ 0.0013 &
 ~.046 & ~.002 & ~.003 & ~.020 & ~.012 & ~.001 & $-$.024 & ~1.00 \\ 
 $\Afbzt$      & 0.0188 $\pm$ 0.0017 &
 ~.035 & ~.001 & ~.002 & ~.013 & $-$.003 & ~.009 & $-$.020 & ~.046 & ~1.00 \\ 
\multicolumn{3}{c}{~}\\[-0.5pc]
\multicolumn{2}{c} {with lepton universality} \\
\hline 
\multicolumn{2}{c|}{$\pzz \chi^2/N_{\rm df}\,=\,36.5/31 $}  & 
   $\MZ$ & $\GZ$ & $\shad$ & $\Rl$ &$\Afbpol$ \\
\hline 
 $\MZ$ [\GeV{}]  & 91.1875$\pm$ 0.0021$\pz$    &
 ~1.00 \\ 
 $\GZ$ [\GeV]  & 2.4952 $\pm$ 0.0023$\pz$    &
 $-$.023  & ~1.00 \\ 
 $\shad$ [nb]  & 41.540 $\pm$ 0.037$\pzz$ &
 $-$.045 & $-$.297 &  ~1.00 \\ 
 $\Rl$         & 20.767 $\pm$ 0.025$\pzz$ &
 ~.033 & ~.004 & ~.183 & ~1.00 \\ 
 $\Afbpol$     & 0.0171 $\pm$ 0.0010   & 
 ~.055 & ~.003 & ~.006 & $-$.056 &  ~1.00 \\ 
\end{tabular} } 
\end{center}
\vskip-1pc\caption {\label{tab:lsafbresult}\em \small Result of the combination
of the four sets of nine pseudo-observables from Table\,\ref{tab:Zparinput}.}
\end{table}

The value of $\chi^2$ per degree of freedom of the combination 
is 32.6/27 and corresponds to a probability of 
21\,\% to find an agreement among the four sets of measurements 
which is worse than the one actually observed. The correlation matrix
of the combined result shows significant correlations of 
$\shad$ with $\GZ$ and $\Ree$, $\Rmu$ and $\Rtau$ and between $\Ree$
and $\Afbze$.

\begin{figure}[hbt]\begin{center}
\mbox{\epsfig{file=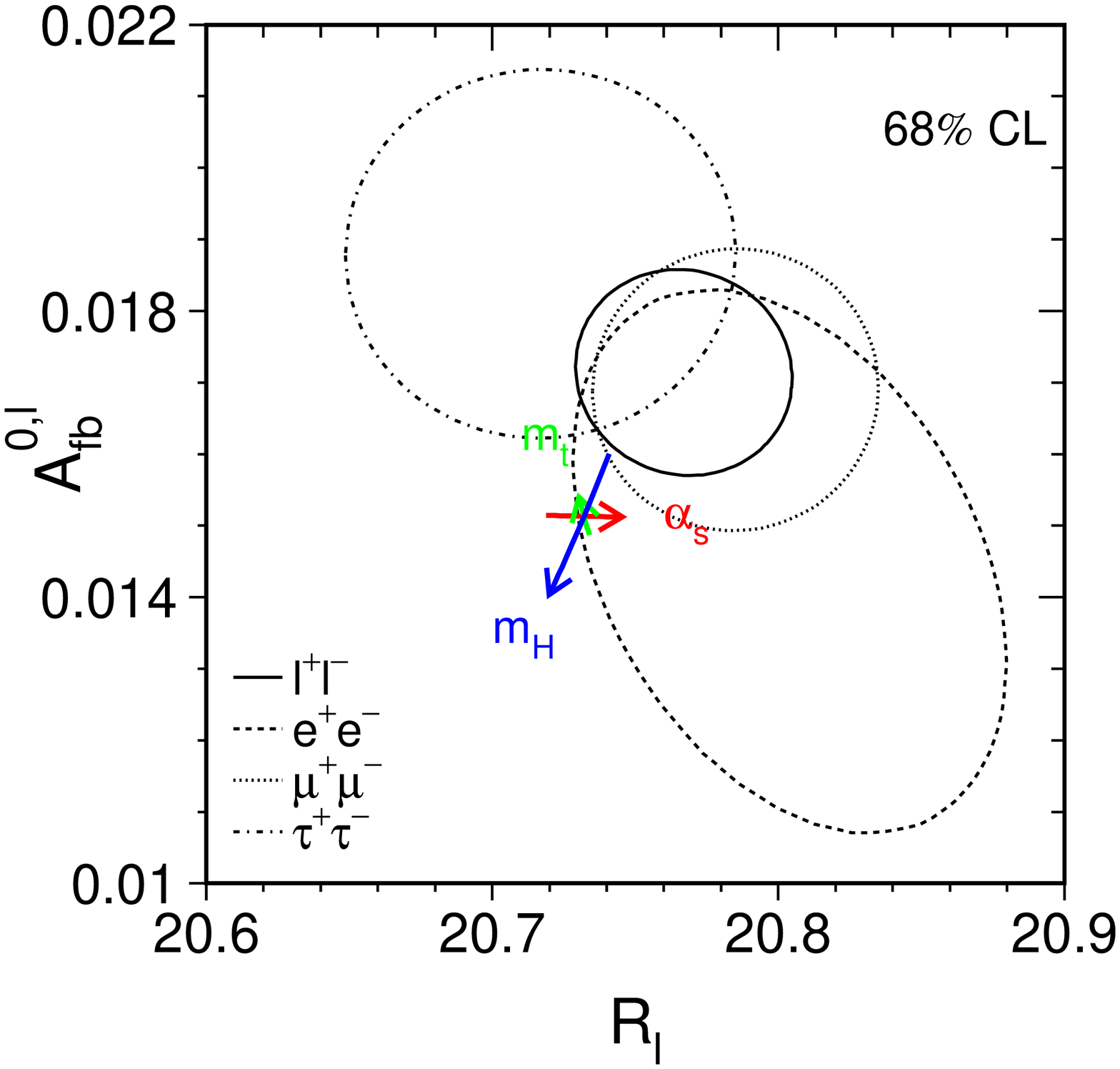,width=0.7\textwidth}}
\end{center}\vspace*{-2pc}
\caption{\label{fig:rlafb} \em \small 
Contour lines (68\,\% CL) in the $\Rl$\,--\,$\Afbpol$ plane for
$\ee$, $\mumu$ and $\tautau$ final states and for all leptons
combined. For better comparison the results for the $\tau$ lepton 
are corrected to correspond to the massless case.  
The SM prediction for $\MZ$=91.1875\,\GeV{}, $\Mt$=174.3\,\GeV{}, 
$\MH$=300\,\GeV{} and $\alpha_s$=0.119 is also shown. 
The lines with arrows correspond to the variation of the SM 
prediction when $\Mt$, $\MH$ and $\alpha_s$ are varied in the 
intervals $\Mt=174.3\pm5.1\,\GeV{}$, 
$\MH=300^{+700}_{-200}\,\GeV{}$, and 
$\alpha_s=0.119\pm0.002$, respectively. 
The arrows point in the direction of increasing values
of $\Mt$, $\MH$ and $\alpha_s$.  }
\end{figure}

A comparison of the leptonic quantites $\Ree$, $\Rmu$ and $\Rtau$
and of $\Afbze$, $\Afbzm$ and $\Afbzt$ shows that they agree within 
errors. Note that $\Rtau$ is expected to be larger 
by 0.23\,\% because of $\tau$ mass effects.
Figure\,\ref{fig:rlafb} shows the corresponding 68\,\% level 
contours in the $\Rl$-$\Afbpol$ plane.

Imposing the additional requirement of lepton universality in the 
combination leads to the results shown in the second part of
Table\,\ref{tab:lsafbresult}. Here $\Rl$ is not a simple average
over the three lepton species, but refers to $\Zzero$ decays 
into pairs of one massless charged lepton species.
The value of $\chi^2/N_{\rm df}$ of 
$36.5/31$ for the combination of the four sets of nine pseudo-observables 
into the five parameters of Table\,\ref{tab:lsafbresult} corresponds to a 
$\chi^2$-probability of 23\,\%. The central ellipse in 
Figure\,\ref{fig:rlafb} shows the 68\,\%-CL contour for the combined 
values of $\Rl$ and $\Afbpol$ determined from all three lepton species.

In principle, the average over the four experiments can also be 
performed at the level of the five parameter results of 
Section\,\ref{sec-fivepar}. When this is attempted, good agreement 
is seen with the results in the last line of 
Table\,\ref{tab:lsafbresult}, except
for $\Rl$, where the difference amounts to $0.005$ or 20\,\% of the 
total error. The origin of this shift is the common $t$-channel error, 
as discussed in Section\,\ref{sec-fivepar} above. 

\subsection{Parameter Transformations \label{sec-partrafo}}
Additional pseudo-observables, more familiar than the experimentally 
motivated set of Table\,\ref{tab:lsafbresult}, can be obtained
by parameter transformations. 
The partial $\Zzero$ decay widths are summarised in 
Table\,\ref{tab:width}; it should be noted that these have larger 
correlations than the original set of results. If lepton 
universality is imposed, the value of $\Ghad$ also changes and 
its error is reduced, because $\Gee$ in the relation between the 
hadronic pole cross-section and the partial widths is replaced 
by the more precise value of $\Gl$. The invisible width, 
$\Ginv=\GZ-\Ghad-\Gee-\Gmumu-\Gtautau$, is also shown in the table.
The leptonic pole cross-section, $\slept$, defined as
\[ \slept\equiv{12\pi\over\MZ^2}{\Gl^2\over\GZ^2} \, , \]
in analogy to $\shad$, is given in the last line of Table\,\ref{tab:width}.
Because QCD final state corrections appear quadratically in the 
denominator via $\GZ$, $\slept$ has a higher sensitivity to $\alpha_s$ 
than $\shad$ or $\Rl$, where the dependence on QCD corrections is 
only linear. 

\begin{table}[hbtp] \begin{center} 
\begin{tabular} {lr|r@{\,}r@{\,}r@{\,}r}
 \multicolumn{2}{c|}{without lepton universality} &
                            \multicolumn{4}{l}{~~~correlations} \\
\hline 
$\Ghad$ [\MeV]     & 1745.8$\pz\pm$2.7$\pz\pzz$ & ~1.00 \\
$\Gee$ [\MeV]      & 83.92$\pm$0.12$\pzz$& $-$0.29 & ~1.00 \\ 
$\Gmumu$ [\MeV]    & 83.99$\pm$0.18$\pzz$& ~0.66 & $-$0.20 & ~1.00 \\
$\Gtautau$ [\MeV]  & 84.08$\pm$0.22$\pzz$&  0.54 & $-$0.17 & ~0.39 & ~1.00 \\  
\hline 
    \multicolumn{6}{c}{~} \\[-0.5pc]
    \multicolumn{2}{c}{with lepton universality} \\
\hline 
$\Ginv$ [\MeV]     & $\pz$499.0$\pzz\pm$1.5$\pz\pzz$    & ~1.00 \\
$\Ghad$ [\MeV]     & 1744.4$\pzz\pm$2.0$\pz\pzz$   & $-$0.29 & ~1.00 \\
$\Gl$ [\MeV]       & 83.984$\pm$0.086$\pz$  & ~0.49 & ~0.39 & ~1.00 \\
\hline 
$\Ginv/\Gl$        & {$\pz$5.942$\pm$0.016$\pz$} &    \\
\hline 
$\slept$ [nb]      & {2.0003$\pm$0.0027} &  \\
\end{tabular} \end{center} 
\caption[Partial widths]{\label{tab:width} \em \small Partial $\Zzero$ decay
widths and correlation coefficients. }
\end{table}

Assuming only standard particles, the invisible width is 
compatible with the SM hypothesis of
decays into the three neutrino species, $\Ginv\,=\,3\,\Gnu$.
The ``number of neutrinos'', $N_\nu$, is calculated according to
\[ \frac{\Ginv}{\Gl} =
 N_\nu \left(\frac{\Gnu}{\Gl}\right)_{\rm SM}. \]
The Standard Model value for the ratio of the partial widths to neutrinos
and to charged leptons is 1.9912$\pm$0.0012, where the uncertainty 
arises from variations of the top quark mass within
its experimental error and of the Higgs mass within
$100\,\GeV\,<\,\MH\,<\,1000\,\GeV$.
With the measured value of $\Ginv\,/\,\Gl = 5.942 \pm 0.016$,
the number of light neutrino species is determined to be
\[ N_{\nu}=2.9841 \pm 0.0083\;.\]
This may also be turned into a quantitative limit on extra, non-standard
contributions to the invisible width, {\sl i.\,e.} not originating 
from $\Ztonunu$, by taking the difference between
the value given in Table\,\ref{tab:width} and the Standard Model expectation
of $\left({\Ginv}\right)_{\rm SM}=501.7^{+0.1}_{-0.9}\,\MeV$. This gives  
$\Delta\Ginv\,=\,-2.7^{+1.7}_{-1.5}\,\MeV$, or expressed as a limit,
$\Delta\Ginv < 2.0$\,\MeV ~@ 95\% CL; here, the limit was
conservatively calculated allowing only positive values of
$\Ginvx$.

The effective axial-vector and vector couplings of the $\Zzero$ 
to leptons, $\gal$ and $\gvl$,  may be expressed in terms of 
the effective Veltman $\rho$ parameter~\cite{RhoVeltman} and the 
effective weak mixing angle, $\swsqeff$, by 
\[ \begin{array}{lcl}
\gal & = &\sqrt{\rhoeffl} I_3^{\ell}\;,        \\
     &   &                              \\[-0.7pc]
\gvl& = &\sqrt{\rhoeffl}(I_3^{\ell}
              -2Q^{\ell}\swsqeffl)\,, \\
\end{array} \]
where $I_3^{\ell}=-\frac{1}{2}$ is the weak iso-spin of 
charged leptons.

The leptonic pole asymmetry, $\Afbpol$, depends only 
on the ratio of the universal lepton couplings $\gvl\,/\,\gal$,
as is easily seen from its definition given in Section\,\ref{sec-zpar},
and thus directly determines the effective weak mixing angle,
$$\swsqeffl\,=\,
  \frac{1}{4}\left(1\,- \frac{\gvl}{\gal} \right)\,,$$ 
if the sign of the ratio of couplings is chosen to be positive, 
in agreement with studies of the polarisation of $\tau$ leptons
at {\Lep}.

The leptonic width is almost entirely given by the parameter $\rhoeffl$, 
with a small contribution from the weak mixing angle entering 
through the vector couplings. A simultaneous fit of $\Gl$ and $\Afbpol$ 
for $\rhoeffl$ and $\swsqeffl$ results in
\[ \begin{array}{ccl}
\rhoeffl  &=& 1.0048\pz \,\pm\, 0.0011\,,\pz \\
          & &                   \\[-0.7pc]
\swsqeffl &=& 0.23099   \,\pm\, 0.00053\,, \\  
\end{array} \]
with an error correlation coefficient of ~27\,\%\,.

Information on the effective $\rho$ parameter contained in $\Ginv$ 
and $\Ghad$ does not significantly improve on this result. 
From $\Ginv$, the effective $\rho$ parameter for neutrinos, 
$\rhoeffnu=(2\,\gnu)^2$, is determined to be 
$\rhoeffnu=1.0027\pm0.0030$. The extraction of $\rho$ from
the hadronic width would require external constraints 
on the strong coupling constant and SM assumptions on 
flavour-dependent corrections. 

The result on $\swsqeffl$ derived here from the leptonic forward-backward
asymmetries agrees well with the value recently published 
by SLD based on measurements of the left-right 
polarisation asymmetry of hadron production at the $\Zzero$ 
resonance\,\cite{bib-ALR}. 
The above value of $\rhoeffl$ is  4.4~standard deviations 
greater than its tree-level value of one, in good agreement with full 
SM calculations, and thus clearly shows the presence of genuine 
electroweak corrections. 

\section{Conclusion \label{sec-conclusion}}

The combination procedure adopted by the LEP electroweak working group
for the four sets of pseudo-observables derived by the {\Lep}
experiments from the measured hadronic 
and leptonic cross-sections and the leptonic forward-backward 
asymmetries at centre-of-mass energies 
around the $\Zzero$ resonance has been described. 

The combination procedure averages parameters extracted from
the measurements of each individual experiment.
This approximates a statistically optimal average that would
be based on the measured cross-sections and forward-backward 
asymmetries to better than 10\,\% of the combined errors, 
{\it i.\,e.} the chosen set of parameters adequately represents the
full set of measurements. The technical precision of the 
adopted combination procedure is around 5\,\% of the combined errors. 
Using each of the four sets of input parameters in the framework of the 
minimal Standard Model yields results which are on average 
identical, to within 5\,\% of the combined errors, to those 
obtained directly from Standard-Model fits to the measured 
cross-sections and asymmetries of each individual experiment.

Detailed studies of common systematic errors were performed. The
dominant contribution of $\pm 1.7\,\MeV$ to the combined error 
on the $\Zzero$ mass arises from the calibration of the energy 
of the beams in {\Lep}. The dominant contribution of $\pm 0.025$\,nb 
to the uncertainty on the hadronic cross-section at the pole of the 
$\Zzero$ resonance arises from the 
theoretical error on the small-angle Bhabha cross-section. The
errors on all other parameters are dominated by independent 
experimental or statistical errors.

The combined {\Lep} results on the mass and width of
the $\Zzero$, on the hadronic pole cross-section, on the ratio
of the hadronic and leptonic partial width and on the pole 
forward-backward asymmetry are 
\vskip -1pc \[ \begin{array}{lcr@{\,}c@{\,}l}
 \MZ     &=& 91.1875&\pm& 0.0021\,\GeV\,,   \\
 \GZ     &=& 2.4952 &\pm& 0.0023\,\GeV\,,   \\
 \shad   &=& 41.540\pz&\pm& 0.037\,{\rm nb}\,, \\
 \Rl     &=& 20.767\pz&\pm& 0.025\,,   \\
 \Afbpol &=& 0.0171   &\pm& 0.0010\,.\\
\end{array} \]

Correlations are typically small ($\leq$ 5\,\%), 
being significant only between $\shad$ and $\GZ$ ($-30\,\%$) and 
between $\shad$ and $\Rl$ ($18\,\%$). The full set of results
and the error correlation matrices are shown in 
Table\,\ref{tab:lsafbresult}.

\pagebreak[1]
\section*{Acknowledgements}
We wish to thank the CERN SL division for 
the excellent performance of the {\Lep} collider, and the working group on 
energy calibration for providing a precise knowledge on the beam energies.  
We would also like to thank all our theorist colleagues who have contributed
to the precision calculations of observables at the $\Zzero$ resonance, 
and the TOPAZ0 and ZFITTER teams who have incorporated these 
calculations and made them available to us. 

\clearpage\newpage

\begin{appendix}
\section{Appendix \label{sec-appendix}}

\subsection{Common errors estimated by individual experiments}

\begin{table}[htbp] \begin{center}
\begin{tabular} {r||l|rrrr}
  ~   &    ~    &   $\MZ\pz$     & $\GZ\pz$      & $\shad\pz$ & $\Ree\pz$ \\
\hline
{\Aleph} & $\MZ$\,[\GeV]  &   0.0017&          &        & \\
 ~    &$\GZ$\,[\GeV]   &$-$0.0004&   0.0012&        & \\
 ~    &$\shad$\,[nb] &$-$0.0028&$-$0.0024& 0.011 & \\
 ~    &$\Ree$  &   0.0013&$-$0.0015& 0.007 & 0.012\\
\hline 
{\Delphi} & $\MZ$\,[\GeV] &   0.0016  & &  & \\
 ~    &$\GZ$\,[\GeV]   &$-$0.0005  &   0.0012 &  & \\
 ~    &$\shad$\,[nb] &$-$0.0025  &$-$0.0024 &  0.009 & \\
 ~    &$\Ree$  &   0.0014&     0.0000 &  0.004 &  0.016 \\ 
\hline 
{\Lthree}    &$\MZ$\,[\GeV]   &   0.0016& &  & \\
 ~    &$\GZ$\,[\GeV]   &$-$0.0008&   0.0013 &  & \\
 ~    &$\shad$\,[nb] &$-$0.0009&$-$0.0030&   0.011& \\
 ~    &$\Ree$  &   0.0020&$-$0.0021&   0.010&  0.013 \\ 
\hline 
{\Opal}  &$\MZ$\,[\GeV]   &   0.0017& & & \\   
 ~    &$\GZ$\,[\GeV]   &$-$0.0005&   0.0012& & \\
 ~    &$\shad$\,[nb] &$-$0.0011&$-$0.0028& 0.011 &\\  
 ~    &$\Ree$  &   0.0019&$-$0.0019& 0.008 &0.012 \\ 
\end{tabular}\end{center}
\caption {\label{tab:cove9ap}\em \small 
Common energy errors, from individual data sets for the 
nine-parameter fits. Values are given as the signed square root of
the covariance matrix elements; elements above the diagonal have been
omitted for simplicity.} \end{table}

\begin{table}[htbp] \begin{center}
\begin{tabular} {r||l|rrr}
  ~   &    ~    & $\Afbze\pz$ & $\Afbzm\pz$ & $\Afbzt\pz$ \\
\hline 
{\Aleph} & $\Afbze$ &   0.0003&          &        \\
  ~   & $\Afbzm$ &$-$0.0003&  0.0003 &        \\
      & $\Afbzt$ &$-$0.0003&  0.0003 & 0.0003 \\
\hline 
{\Delphi}& $\Afbze$ &   0.0004&         &        \\
  ~   & $\Afbzm$ &$-$0.0003&  0.0003 &        \\
      & $\Afbzt$ &$-$0.0003&  0.0003 & 0.0003 \\
\hline 
{\Lthree}    & $\Afbze$ &   0.0003&         &        \\
  ~   & $\Afbzm$ &$-$0.0002&  0.0003 &        \\
      & $\Afbzt$ &$-$0.0002&  0.0002 & 0.0003 \\
\hline 
{\Opal}  & $\Afbze$ &   0.0004&          &        \\
  ~   & $\Afbzm$ &$-$0.0003&  0.0003 &        \\
      & $\Afbzt$ &$-$0.0003&  0.0003 & 0.0003 \\
\end{tabular}
\end{center}
\caption {\label{tab:cove_afbap}\em \small
 Common energy errors for forward-backward asymmetries,
estimated from the data sets of individual experiments.}
\end{table} 

\begin{table}[htbp] \begin{center}
\begin{tabular} {r||l|rr}
  ~   &    ~    & $\Ree\pzz$ & $\Afbze\pz$ \\
\hline 
{\Aleph} & $\Ree$ &   0.025$\pz$&            \\
  ~   & $\Afbze$ & $-$0.0056&  0.0013 \\
\hline 
{\Delphi} & $\Ree$ &   0.025$\pz$&            \\
  ~   & $\Afbze$ &$-$0.0058&  0.0016 \\
\hline 
{\Lthree}    & $\Ree$   &   0.021$\pz$ &            \\
  ~   & $\Afbze$ &$-$0.0046&  0.0010 \\
\hline 
{\Opal} & $\Ree$ &   0.025$\pz$&            \\
  ~   & $\Afbze$ &$-$0.0057&  0.0015 \\
\end{tabular}
\end{center}
\caption {\label{tab:tch_systap}\em \small
 Estimates of the $t$-channel errors of individual experiments.} 
\end{table} 

\clearpage

\subsection{Results with lepton universality}

\begin{table}[htbp] \begin{center}{\small
\begin {tabular} {lr|r@{\,}r@{\,}r@{\,}r@{\,}r@{\,}}
\multicolumn{2}{c|}{~}& \multicolumn{5}{c}{correlations} \\
\multicolumn{2}{c|}{~} & $\MZ$ & $\GZ$ & $\shad$ & $\Rl$ & $\Afbpol$ \\
\hline 
\multicolumn{2}{l}{$\pzz \chi^2/N_{\rm df}\,=\,172/180$}     & 
                            \multicolumn{1}{c}{\Aleph} \\
\hline 
 $\MZ$\,[\GeV]\hspace*{-.5pc} & 91.1893 $\pm$ 0.0031     &   
 1.00 \\
 $\GZ$\,[\GeV]\hspace*{-2pc}  &  2.4959 $\pm$ 0.0043     & 
 .038 & 1.00 \\ 
 $\shad$\,[nb]\hspace*{-2pc}  &  41.559 $\pm$ 0.057$\pz$ &   
 $-$.092 & $-$.383 & 1.00 \\ 
 $\Rl$      &  20.729 $\pm$ 0.039$\pz$ &   
 .033 & .011 & .246 & 1.00 \\
 $\Afbpol$     &  0.0173 $\pm$ 0.0016     &   
 .071  & .002 & .001 & $-$.076 & 1.00 \\
    ~          & \multicolumn{2}{c}{~}                \\[-0.2pc]
\hline 
\multicolumn{2}{l}{$\pzz \chi^2/N_{\rm df}\,=\,183/172 $} & 
                             \multicolumn{1}{c}{\Delphi} \\
\hline 
 $\MZ$\,[\GeV]\hspace*{-.5pc}   &  91.1863 $\pm$ 0.0028    &
 1.00 \\
 $\GZ$\,[\GeV]\hspace*{-2pc}   &  2.4876 $\pm$ 0.0041     &
 .046 & 1.00 \\
 $\shad$\,[nb]\hspace*{-2pc}   &  41.578 $\pm$ 0.069$\pz$ &
 $-$.070 & $-$.270  &1.00 \\
 $\Rl$      &  20.730  $\pm$ 0.060$\pz$ &
 .028 & $-$.006 & .242 & 1.00 \\
 $\Afbpol$     &  0.0187 $\pm$ 0.0019     &
 .095 & .006 & $-$.005 & .000 & 1.00 \\
    ~          & \multicolumn{2}{c}{~}                \\[-0.2pc]
\hline 
\multicolumn{2}{l}{$\pzz \chi^2/N_{\rm df}\,=\,163/170$}  & 
                                     \multicolumn{1}{c}{\Lthree} \\
\hline 
 $\MZ$\,[\GeV]\hspace*{-.5pc}   &  91.1894 $\pm$ 0.0030      & 
 1.00 \\
 $\GZ$\,[\GeV]\hspace*{-2pc}   &   2.5025 $\pm$ 0.0041      & 
 .068 & 1.00 \\
 $\shad$\,[nb]\hspace*{-2pc}   &   41.536 $\pm$ 0.055$\pz$  & 
 .014 & $-$.348 & 1.00 \\
 $\Rl$      &   20.809  $\pm$ 0.060$\pz$ & 
 .067 & .020 & .111 & 1.00 \\
 $\Afbpol$     &   0.0192 $\pm$ 0.0024      & 
 .041 & .020 & .005 & $-$.024 & 1.00 \\
    ~          & \multicolumn{2}{c}{~}                \\[-0.2pc]
\hline 
\multicolumn{2}{l}{$\pzz \chi^2/N_{\rm df}\,=\,158/198$} 
                                      & \multicolumn{1}{c}{\Opal} \\
\hline 
 $\MZ$\,[\GeV]\hspace*{-.5pc}   & 91.1853 $\pm$ 0.0029       &
 1.00 \\
 $\GZ$\,[\GeV]\hspace*{-2pc}   & 2.4947 $\pm$ 0.0041        &
 .051 & 1.00 \\
 $\shad$\,[nb]\hspace*{-2pc}   & 41.502 $\pm$ 0.055$\pz$    &
 .030 & $-$.352 & 1.00 \\
 $\Rl$      & 20.822 $\pm$ 0.044$\pz$    &
 .043 & .024 & .290 & 1.00 \\
 $\Afbpol$     & 0.0145 $\pm$ 0.0017        &
 .075 & $-$.005 & .013 & $-$.017 & 1.00 \\ 
\end{tabular}}
\end{center}
\caption[Five parameter results]{\label{tab:Zpar5input} \em \small
Results on Z parameters and error correlation matrices by the four
experiments with lepton universality imposed.}
\end{table} 

\clearpage
\newpage

\subsection{Results from eleven-parameter fits\label{sec-Zpar11input}}

\begin{tabular}{r|r}
\multicolumn{2}{c}{\Aleph} \\ 
\hline 
 $\GZ$ [\GeV] &   2.4957 $\pm$ 0.0043  \\  
 $\shad$ [nb] &  41.559$\pz$  $\pm$  0.057$\pz$  \\
 $\Ree$       &  20.694$\pz$  $\pm$  0.075$\pz$  \\
 $\Rmu$       &  20.801$\pz$  $\pm$  0.056$\pz$  \\
 $\Rtau$      &  20.709$\pz$  $\pm$  0.062$\pz$  \\  
 $\Afbze$     &   0.0184 $\pm$  0.0034 \\
 $\Afbzm$     &   0.0173 $\pm$  0.0025 \\   
 $\Afbzt$     &   0.0171 $\pm$  0.0028 \\   
 $\MZA$ [\GeV]&  91.1928 $\pm$  0.0092 \\ 
 $\MZB$ [\GeV]&  91.1926 $\pm$  0.0046 \\ 
 $\MZC$ [\GeV]&  91.1852 $\pm$  0.0043 \\ 
\end{tabular} \\

\hspace*{15em}                  correlation matrix
\[\begin {array} 
 {r@{\,}r@{\,}r@{\,}r@{\,}r@{\,}r@{\,}r@{\,}r@{\,}r@{\,}r@{\,}r@{\,}}
 1.000&-0.384 &0.001 &0.012 &0.003 &0.001 &0.000 &0.000 &0.015&-0.011 &0.055\\
-0.384 &1.000 &0.137 &0.167 &0.152&-0.004 &0.003 &0.003&-0.043&-0.037&-0.074\\
 0.001 &0.137 &1.000 &0.083 &0.067&-0.389 &0.020 &0.017 &0.063 &0.063 &0.036\\
 0.012 &0.167 &0.083 &1.000 &0.093 &0.000 &0.014 &0.000 &0.003&-0.001&-0.004\\
 0.003 &0.152 &0.067 &0.093 &1.000 &0.000 &0.000 &0.012 &0.001&-0.001&-0.004\\
 0.001&-0.004&-0.389 &0.000 &0.000 &1.000&-0.009&-0.008&-0.037&-0.036&-0.020\\
 0.000 &0.003 &0.020 &0.014 &0.000&-0.009 &1.000 &0.018 &0.056 &0.063 &0.022\\
 0.000 &0.003 &0.017 &0.000 &0.012&-0.008 &0.018 &1.000 &0.045 &0.060 &0.018\\
 0.015&-0.043 &0.063 &0.003 &0.001&-0.037 &0.056 &0.045 &1.000 &0.016 &0.011\\
-0.011&-0.037 &0.063&-0.001&-0.001&-0.036 &0.063 &0.060 &0.016 &1.000 &0.093\\
 0.055&-0.074 &0.036&-0.004&-0.004&-0.020 &0.022 &0.018 &0.011 &0.093 &1.000\\
\end{array}\]
~\\

\begin{tabular}{r|r}
\multicolumn{2}{c}{\Delphi} \\ 
\hline 
 $\GZ$ [\GeV] &   2.4878     $\pm$  0.0041 \\
 $\shad$ [nb] &  41.573$\pz$  $\pm$  0.069$\pz$ \\
 $\Ree$       &  20.88$\pzz$ $\pm$  0.12$\pzz$ \\
 $\Rmu$       &  20.650$\pz$ $\pm$  0.076$\pz$ \\
 $\Rtau$      &  20.84$\pzz$ $\pm$  0.13$\pzz$ \\
 $\Afbze$     &  0.0170  $\pm$      0.0049 \\
 $\Afbzm$     &  0.0164  $\pm$      0.0025 \\
 $\Afbzt$     &  0.0240  $\pm$      0.0037 \\
 $\MZA$ [\GeV]& 91.1883  $\pm$ 0.0084 \\
 $\MZB$ [\GeV]& 91.1824  $\pm$ 0.0043 \\
 $\MZC$ [\GeV]& 91.1894  $\pm$ 0.0038 \\
\end{tabular}
\clearpage

\hspace*{15em}                  correlation matrix
\[\begin {array} 
 {r@{\,}r@{\,}r@{\,}r@{\,}r@{\,}r@{\,}r@{\,}r@{\,}r@{\,}r@{\,}r@{\,}}
 1.000&-0.271&-0.001&-0.007&-0.001&-0.001 &0.005 &0.003 &0.002 &0.012 &0.062\\
-0.271 &1.000 &0.123 &0.191 &0.113&-0.003 &0.002 &0.001&-0.003 &0.015&-0.066\\
-0.001 &0.123 &1.000 &0.054 &0.033&-0.105 &0.027 &0.016 &0.055 &0.044 &0.029\\
-0.007 &0.191 &0.054 &1.000 &0.051 &0.000 &0.008 &0.000 &0.000&-0.002&-0.002\\
-0.001 &0.113 &0.033 &0.051 &1.000&-0.001 &0.000 &0.011&-0.001&-0.001 &0.002\\
-0.001&-0.003&-0.105 &0.000&-0.001 &1.000&-0.014&-0.014 &0.048 &0.052 &0.016\\
 0.005 &0.002 &0.027 &0.008 &0.000&-0.014 &1.000 &0.015 &0.057 &0.053 &0.021\\
 0.003 &0.001 &0.016 &0.000 &0.011&-0.014 &0.015 &1.000 &0.032 &0.037 &0.016\\
 0.002&-0.003 &0.055 &0.000&-0.001 &0.048 &0.057 &0.032 &1.000 &0.016 &0.007\\
 0.012 &0.015 &0.044&-0.002&-0.001 &0.052 &0.053 &0.037 &0.016 &1.000 &0.088\\ 
 0.062&-0.066 &0.029&-0.002 &0.002 &0.016 &0.021 &0.016 &0.007 &0.088 &1.000\\
\end{array}\]
~\\

\begin{tabular}{r|r}
\multicolumn{2}{c}{\Lthree} \\
\hline 
 $\GZ$ [\GeV] & 2.5024  $\pm$ 0.0041 \\
 $\shad$ [nb] & 41.540$\pz$  $\pm$ 0.055$\pz$ \\
 $\Ree$       & 20.821$\pz$ $\pm$ 0.089$\pz$ \\
 $\Rmu$       & 20.861$\pz$ $\pm$ 0.097$\pz$ \\
 $\Rtau$      & 20.79$\pzz$ $\pm$ 0.13$\pzz$ \\
 $\Afbze$     & 0.0106  $\pm$ 0.0058 \\
 $\Afbzm$     & 0.0190  $\pm$ 0.0033 \\
 $\Afbzt$     & 0.0261  $\pm$ 0.0047 \\
 $\MZA$ [\GeV]& 91.1973 $\pm$ 0.0092 \\
 $\MZB$ [\GeV]& 91.1912 $\pm$ 0.0047 \\ 
 $\MZC$ [\GeV]& 91.1870 $\pm$ 0.0041 \\
\end{tabular}\\

\hspace*{15em}                  correlation matrix
\[\begin {array} 
 {r@{\,}r@{\,}r@{\,}r@{\,}r@{\,}r@{\,}r@{\,}r@{\,}r@{\,}r@{\,}r@{\,}}
 1.000&-0.354&-0.001&-0.001 &0.004&-0.002&-0.001 &0.000 &0.017 &0.014 &0.072 \\
-0.354 &1.000 &0.081 &0.078 &0.055 &0.011 &0.011 &0.007 &0.087 &0.022&-0.038 \\
-0.001 &0.081 &1.000 &0.030 &0.024&-0.150 &0.020 &0.014 &0.117 &0.050 &0.059 \\
-0.001 &0.078 &0.030 &1.000 &0.020&-0.002 &0.005&-0.000 &0.002&-0.001&-0.001 \\
 0.004 &0.055 &0.024 &0.020 &1.000&-0.004&-0.000 &0.009 &0.002 &0.001 &0.002 \\
-0.002 &0.011&-0.150&-0.002&-0.004 &1.000 &0.011&-0.007&-0.044&-0.016&-0.037 \\
-0.001 &0.011 &0.020 &0.005&-0.000 &0.011 &1.000 &0.007 &0.051 &0.047 &0.010 \\
 0.000 &0.007 &0.014&-0.000 &0.009&-0.007 &0.007 &1.000 &0.025 &0.034 &0.008 \\
 0.017 &0.087 &0.117 &0.002 &0.002&-0.044 &0.051 &0.025 &1.000 &0.004 &0.002 \\
 0.014 &0.022 &0.050&-0.001 &0.001&-0.016 &0.047 &0.034 &0.004 &1.000 &0.083 \\
 0.072&-0.038 &0.059&-0.001 &0.002&-0.037 &0.010 &0.008 &0.002 &0.083 &1.000 \\
\end{array}\]
~\\

\begin{tabular}{r|r}
\multicolumn{2}{c}{\Opal} \\
\hline 
 $\GZ$ [\GeV] &  2.4946   $\pm$ 0.0042 \\
 $\shad$ [nb] &  41.506$\pz$   $\pm$ 0.057$\pz$ \\
 $\Ree$       &  20.903$\pz$   $\pm$ 0.085$\pz$ \\
 $\Rmu$       &  20.812$\pz$   $\pm$ 0.058$\pz$ \\
 $\Rtau$      &  20.833$\pz$   $\pm$ 0.091$\pz$ \\
 $\Afbze$     &  0.0089   $\pm$ 0.0045 \\
 $\Afbzm$     &  0.0159   $\pm$ 0.0023 \\
 $\Afbzt$     &  0.0146   $\pm$ 0.0030 \\
 $\MZA$ [\GeV]&  91.1851  $\pm$ 0.0091 \\
 $\MZB$ [\GeV]&  91.1872  $\pm$ 0.0046 \\
 $\MZC$ [\GeV]&  91.1849  $\pm$ 0.0039 \\
\end{tabular}\\

\hspace*{15em}                  correlation matrix
\[\begin {array} 
 {r@{\,}r@{\,}r@{\,}r@{\,}r@{\,}r@{\,}r@{\,}r@{\,}r@{\,}r@{\,}r@{\,}}
 1.000&-0.369 &0.001 &0.019 &0.012&-0.002&-0.009&-0.007 &0.002&-0.065 &0.107 \\
-0.369 &1.000 &0.164 &0.218 &0.137 &0.005 &0.018 &0.014 &0.009 &0.185&-0.100 \\
 0.001 &0.164 &1.000 &0.090 &0.042&-0.228 &0.034 &0.026 &0.079 &0.106 &0.032 \\
 0.019 &0.218 &0.090 &1.000 &0.058&-0.002 &0.012&-0.001&-0.007 &0.005 &0.000 \\
 0.012 &0.137 &0.042 &0.058 &1.000 &0.000 &0.000 &0.016&-0.007 &0.007&-0.001 \\
-0.002 &0.005&-0.228&-0.002 &0.000 &1.000&-0.017&-0.013&-0.040&-0.046&-0.021 \\
-0.009 &0.018 &0.034 &0.012 &0.000&-0.017 &1.000 &0.019 &0.062 &0.073 &0.024 \\
-0.007 &0.014 &0.026&-0.001 &0.016&-0.013 &0.019 &1.000 &0.045 &0.058 &0.019 \\
 0.002 &0.009 &0.079&-0.007&-0.007&-0.040 &0.062 &0.045 &1.000 &0.021 &0.006 \\
-0.065 &0.185 &0.106 &0.005 &0.007&-0.046 &0.073 &0.058 &0.021 &1.000 &0.093 \\
 0.107&-0.100 &0.032 &0.000&-0.001&-0.021 &0.024 &0.019 &0.006 &0.093 &1.000 \\
\end{array}\]\\ ~\\

The common energy errors are shown in Table\,\ref{tab:cove11}.
The agreement of the energy-related errors for the eleven parameter 
results is acceptable among the experiments. Noticeable are the
differences in the covariance matrix elements between the $\Zzero$ mass
values and $\shad$; they were traced back to slightly different 
strategies in choosing a value of $\MZ$ in the different time periods 
for calculations involving the hadronic pole cross section.

\begin{table}[hbtp] \begin{center}
\begin{tabular} {r||l|rrrrrr}
  ~   &   ~      &   $\GZ$  & $\shad$&$\Ree$ & $\MZA$ & $\MZB$ & $\MZC$ \\
\hline
{\Aleph} 
 ~    &$\MZA$  &$-$0.0011 &$-$0.002& 0.0020& 0.0058 &        &        \\
 ~    &$\MZB$  &$-$0.0007 &$-$0.003& 0.0016& 0.0006 & 0.0028 &        \\
 ~    &$\MZC$  &   0.0002 &$-$0.002& 0.0003& 0.0003 & 0.0013 & 0.0015 \\
\hline 
{\Delphi}
 ~    &$\MZA$ &$-$0.0009 &    0.0014& 0.0027 &0.0053  &  &  \\
 ~    &$\MZB$ &$-$0.0006 & $-$0.0020& 0.0014 &0.0008  &0.0027  &  \\
 ~    &$\MZC$ &$-$0.0002 & $-$0.0018& 0.0007 &0.0004  &0.0012  &0.0014  \\
\hline 
{\Lthree}    
 ~    &$\MZA$ &$-$0.0013 &   0.0045&   0.0046&  0.0050  &  &  \\
 ~    &$\MZB$ &$-$0.0007 &$-$0.0022&$-$0.0007&$-$0.0008  & 0.0028 &   \\
 ~    &$\MZC$ &$-$0.0001 &$-$0.0018&$-$0.0006&$-$0.0009  & 0.0012 & 0.0015 \\
\hline 
{\Opal}  
 ~    &$\MZA$ &$-$0.0011&   0.0025& 0.0036&   0.0058 &  & \\
 ~    &$\MZB$ &$-$0.0010&   0.0034& 0.0034&   0.0006 & 0.0030 & \\
 ~    &$\MZC$ &   0.0000&$-$0.0014& 0.0013&$-$0.0003 & 0.0012 & 0.0015 \\
\hline 
\hline 
average 
 ~    &$\MZA$ &$-$0.0011&   0.0016&   0.0032&   0.0055&  &  \\
 ~    &$\MZB$ &$-$0.0008&$-$0.0010&   0.0014&   0.0003& 0.0028&  \\
 ~    &$\MZC$ &   0.0000&$-$0.0018&$-$0.0004&$-$0.0001& 0.0012& 0.0015 \\
\end{tabular}
\end{center}
\vspace*{-0.5pc} 
\caption {\label{tab:cove11}\em \small 
Common energy errors for multiple-$\MZ$ fit, from 
individual data sets and average. The matrix elements for $\GZ$, $\shad$ and
$\Ree$ and for the asymmetries are very similar to the nine-parameter case
and can be taken from Table\,\ref{tab:cove9ap} and\,\ref{tab:cove_afbap}.}
\end{table}
\clearpage
\end{appendix}
\clearpage\newpage

\bibliographystyle{unsrt}
\bibliography{zpar}
%
\end{document}